## Spin-driven electrical power generation at room temperature

K. Katcko[1], E. Urbain[1], B. Taudul[1], F. Schleicher[1,2], J. Arabski[1], E. Beaurepaire[1†], B. Vileno[3], D. Spor[1], W. Weber[1], D. Lacour[2], S. Boukari[1], M. Hehn[2], M. Alouani[1], J. Fransson[4], M. Bowen[1*]

*[1] Institut de Physique et Chimie des Matériaux de Strasbourg, UMR 7504 CNRS, Université de Strasbourg, 23 rue du Lœss, BP 43, 67034 Strasbourg, France.*

*[2] Institut Jean Lamour UMR 7198 CNRS, Université  de Lorraine, BP 70239, 54506 Vandœuvre lès Nancy, France.*

*[3] Institut de Chimie, UMR 7177 CNRS, Université de Strasbourg, 4 rue Blaise Pascal, CS 90032, 67081 Strasbourg, France.*

*[4] Department of Physics and Astronomy, Uppsala University, Box 516, 75120, Uppsala, Sweden.*

[†]: deceased April 24th, 2018.

* email: bowen@unistra.fr

To mitigate climate change, our global society is harnessing direct (solar irradiation) and indirect (wind/water flow) sources of renewable electrical power generation. Emerging direct sources include current-producing thermal gradients in thermoelectric materials, while quantum physics-driven processes to convert quantum information into energy have been demonstrated at very low temperatures. The magnetic state of matter, assembled by ordering the electron's quantum spin property, represents a sizeable source of built-in energy. We propose to create a direct source of electrical power at room temperature (RT) by utilizing magnetic energy to harvest thermal fluctuations on paramagnetic (PM) centers. Our spin engine rectifies current fluctuations across the PM centers' spin states according to the electron spin by utilizing so-called 'spinterfaces' with high spin polarization. As a rare experimental event, we demonstrate how this path can generate 0.1nW at room temperature across a 20μm-wide spintronic device called the magnetic tunnel junction, assembled using commonplace Co, C and MgO materials. The presence of this path in our experiment, which also generates very high spintronic performance, is confirmed by analytical and ab-initio calculations. Device downscaling, and the ability for other materials systems than the spinterface to select a transport spin channel at RT widens opportunities for routine device reproduction. The challenging control over PM centers within the tunnel barrier's nanotransport path may be addressed using oxide- and organic-based nanojunctions. At present densities in MRAM products, this spin engine could lead to 'always-on' areal power densities well beyond that generated by solar irradiation on earth. Further developing this concept can fundamentally alter our energy-driven society's global economic, social and geopolitical constructs.

**Main Text**

A solar cell's electronic potential landscape is crafted such that, when a photon is absorbed, the resulting electron and hole (the absence of an electron) flow in opposite directions. Since they carry an electrical charge of opposite sign, this generates an electrical current. Two low-temperature experiments[1,2] have suggested that, by astutely designing the magnetic potential landscape of a quantum dot (QD) device, electrons with a spin ↑ or ↓ quantum property can flow in opposite directions. This can generate electrical power if the spin ↑ and ↓ current channels are imbalanced, i.e. if the overall current is spin-polarized. This apparent current imbalance, and the presence of QDs in both systems, are reminiscent of quantum thermodynamical experiments on single electron boxes, which have demonstrated how to harvest thermal fluctuations[3,4] and information[5] to perform work at very low temperatures. These heat and information engines are driven by fluctuation-induced quantum tunnelling on/off of QDs, with a transmission asymmetry between left and right leads that can be energy-dependent due to the QD's discrete energy levels[6,7]. A few reports have theoretically[8] and experimentally (using NV centers[9]) taken into account the electron spin.

Inspired by the report of Miao et al[2], and by recent progress in quantum thermodynamics[3,5–7], we propose that a spin-split paramagnetic (PM) quantum object can enable electrons with a spin ↑ or ↓ quantum property to flow in opposite directions if the transmission rates on either side of the PM center are spin-dependent. Fig. 1(a) illustrates the simplified case of a PM center, characterized by two effectively spin-split energy levels. To achieve a strongly spin-dependent transmission rate $\Gamma$, the PM center is placed between spintronic selectors --- materials systems that ideally favor only one transport spin channel while blocking the other. Examples include half metals[10–12], 2D materials with half-metallic properties[13], a normal metal / ferromagnetic tunnel barrier[14] bilayer, and the ferromagnetic metal/molecule interface[15] (aka 'spinterface', see hereafter). The Fe/MgO system may also constitute a spintronic selector given either a sufficient MgO thickness[16]  and/or the presence of oxygen vacancies[17,18]. Due to this combination of spintronic selectors and spin-split PM states, a spin ↑(↓) electron may only depart the PM toward the left(right) electrode at the energy of the PM center's corresponding spin state.

Hai et al[1] used a MnAs ferromagnetic metal with a conventional (~50%) spin polarization of conduction states, and applied a magnetic field to spin-split their MnAs QDs and obtain power generation. It is unclear whether this experiment could have worked at higher temperatures. Miao et al[2] filtered the electron spin upon transport across EuS ferromagnetic tunnel barriers below its ordering temperature $T_C \sim 16.8K$. Here, spin splitting of the Al Qd is induced by electronic coupling to one EuS barrier.

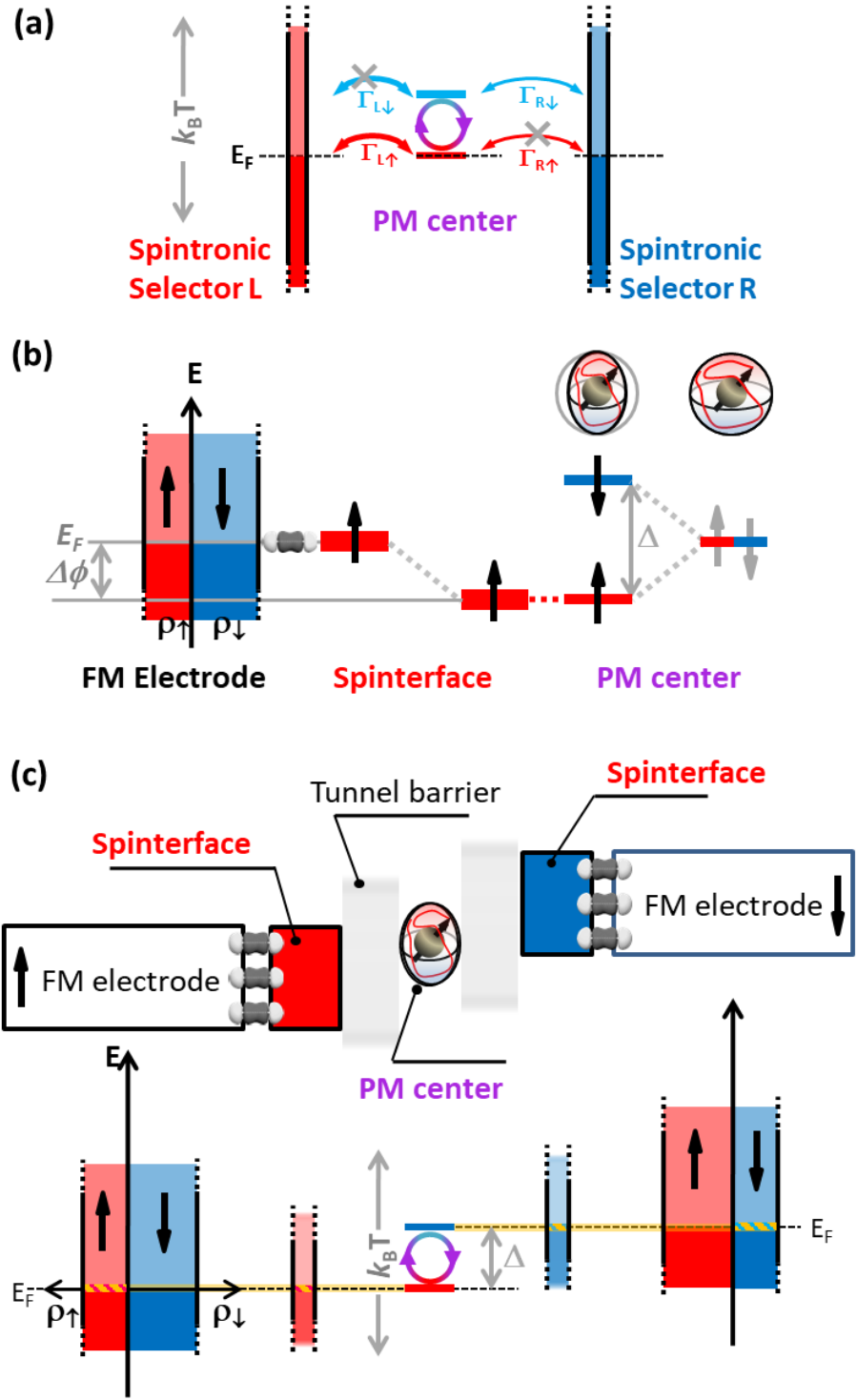

Figure 1: **A room-temperature spin engine.** (a) Illustration of asymmetries in the spin-induced transmission rates Γ across a single PM center between spintronic selectors L and R. See text for possible candidates, including the spinterface. (b) Spin-conserved quantum tunneling between a spinterface and a PM center deforms the PM center's Bloch sphere, thereby splitting[19,20] its spin states by Δ, and shifts the spinterface's Fermi level by Δϕ. (c) The spintronic landscape across a MTJ, comprising spinterfaces and a PM center, in its AP magnetic state exhibits a spontaneous bias voltage ΔV≤Δ. A yellow band designates the thermally rectified, spin-polarized current involving striped segments of the FM and spinterface DOS. The spin engine generates work by harvesting thermal spin fluctuations on the PM center. See text for details. In all panels, red(blue) correspond to spin ↑(↓) states.

Reports indicate that several spintronic selector tracks include materials science candidates (e.g. the Fe/MgO MTJ class[16,18], the half-metallic $Co_2FeAl$[12] or the ferromagnetic tunnel barrier $CoFe_2O_4$[14]) that can operate at/beyond room temperature (RT). To obtain RT electrical generation, and in the process demonstrate it to be a RT spintronic selector, we utilize the spinterface[15,21–25]. This refers to a low energy bandwidth, low density of highly spin-polarized states that arise at room temperature from spin-polarized hybridization between the highly degenerate electronic states of a FM metal such as Co and the few, energetically discrete states of molecules, including carbon atoms[24]. The spinterface is weakly conducting, and its magnetic orientation naturally follows that of the FM metal. So far, only spin-polarized photoemission spectroscopy[24,25] suggests that the spinterface may be a spintronic selector at RT.

We now utilize the case of the spinterface to illustrate several key considerations of how spintronic selectors and PM centers can interact to form the spin engine's transport path. Upon connecting the spinterface to the PM state (see Fig. 1(a)), spin-conserved quantum tunneling conditions the resulting spin-polarized landscape in the following significant ways. First, the spinterface's density of states (DOS) with a spatially oriented spin polarization generates a corresponding spintronic anisotropy in the PM state's stochastic spin distribution[19,20], thereby deforming the PM's Bloch sphere of spin states. This generates an energy difference $\Delta$ between the PM center's spin states, and increases the probability that an electron tunnel onto/off of the PM if its spin is aligned to the spinterface's spin referential. The ensuing preferential charge flow for that spin channel effectively shifts[2] the spinterface/FM metal's Fermi level by $\Delta\phi$ toward that spin state of the PM center. We are thus describing how the spinterface can modify a metal's properties[26], namely its Fermi level position, through an additional mechanism.

In a perfectly symmetric magnetic tunnel junction (MTJ) comprising spin-conserved tunneling between these two key ingredients --- spinterfaces and a PM site ---, no net current I should flow in the MTJ's parallel (P) orientation of electrode magnetizations. However, in the MTJ's antiparallel (AP) magnetic state (see Fig. 1(a)), the two FM electrode Fermi levels are shifted away from one another, each toward the corresponding spinterface-selected spin state of the PM center. The resulting spontaneous bias voltage $\Delta V$ between the FM electrodes thus scales with the amplitude of the spinterface's spin polarization and the energy difference $\Delta$ between the PM center's spin $\uparrow$ and $\downarrow$ states. Since an experimental MTJ cannot be exactly symmetric, one may also anticipate a spontaneous bias, or current, in the MTJ's P state, albeit of lower amplitude.

To generate work, the spin engine harvests energy from the spin fluctuations that are thermally induced on the PM center. This thermal spin state mixing on the PM center enables current to flow from one spinterface to the other, even against the built-in $\Delta V$ in the MTJ's AP magnetic state. The spin engine thus requires that $\Delta \leq k_BT$, and thus a balance between the tunnelling-induced energy shift $\Delta\phi$ of the spinterface state to the PM center's spin state and thermal fluctuations, as weighed by the spinterface's spin polarization. This thermal energy harvesting can be expected to cool the PM center. Furthermore, the fully spin-polarized current flowing across the spinterface perturbs the FM ground state of the electrode through a spin accumulation-induced interfacial resistance[27]. The resulting heat generation must be dissipated for our spin engine to work. Finally, the spinterface's low density of highly spin-polarized states may be beneficial to RT operation. Indeed, it protects the energetically discrete PM spin states against thermal broadening from the FM electrodes. As discussed theoretically in the SI, the thermal fluctuations in current are rectified first upon transport from the FM electrode onto the spinterface, and furthermore upon transport from the spinterface onto the PM center's spin state. This, along with the spinterface's high spin polarization, strongly dampens any energy smearing of the PM center's discrete spin states. The resulting energetically sharp, spin-polarized effective current path involving the striped DOS of the FM electrodes and spinterfaces is schematized in Fig. 1(b) by the yellow band.

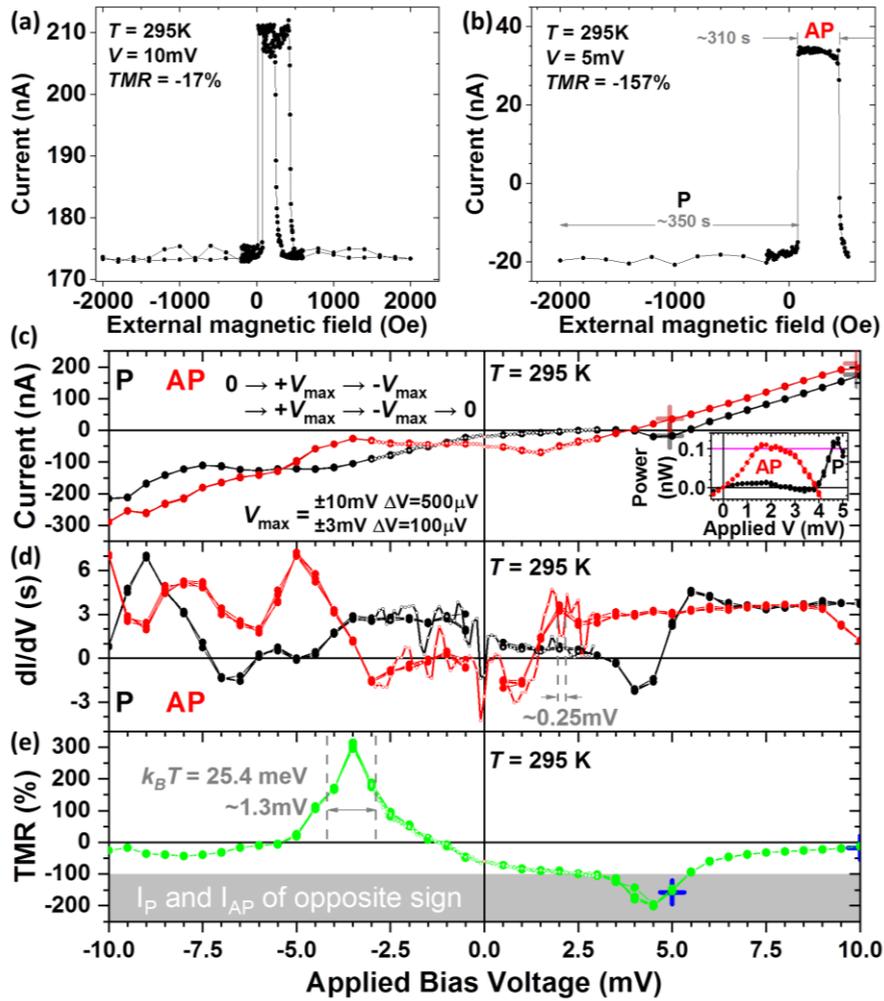

Figure 2: **Spintronics-driven power generation at room temperature.** I(H) curves at (a) +10mV and (b) +5mV measured on the 20μm-diameter MTJ. Bias dependence of (c) current and (d) numerically derived conductance dI/dV in the MTJ's P (black; H=-2000Oe) and AP (red; H=250Oe) states, and (e) the inferred TMR bias dependence. Two sets of 0→+$V_{max}$→-$V_{max}$→+$V_{max}$→-$V_{max}$→0 sweeps spanning ±10mV(±3mV) with 500μV(100μV) steps are shown. Discrete I(V) points obtained from the I(H) data of panels (a) and (b) are represented using semi-transparent crosses. TMR data <-100% (within the gray zone of panel (c); see panel (b)) is achieved when $I_P$ and $I_{AP}$ are of opposite sign at constant V. A non-zero current that depends on the MTJ's magnetic state is observed for V = 0. Despite the ~50meV expected thermal smearing for this 295K measurement, the +300% TMR peak has a FWHM of ~1.3meV and statistically relevant conductance oscillations with a ~0.25meV width are observed (using averaged dI/dV data; see Fig. S4 for full dataset and error bars). This spectral sharpness is a direct signature of the spin engine. Using data from Fig. S3a, the inset to panel (c) shows how the bias-dependent output power depends on the MTJ's magnetic state and can exceed 0.1nW at RT.

As a rare experimental event, we have observed how a MTJ can generate electrical power at room temperature (RT) thanks to our conceptual spin engine, according to analytical and ab-initio theories. As described hereafter, this MTJ integrates Co/C spinterfaces with nearly total spin polarization[24], and paramagnetic C atoms[28] on the oxygen vacancy sites of the MgO tunnel barrier. Referring to Fig. 2a, we observe a negative tunneling magnetoresistance ratio, i.e. TMR=$I_P/I_{AP}$-1 < 0 , at V=+10mV and T = 295K through P/AP magnetic states that are well controlled thanks to an IrMn pinning layer (see Methods). Fig. 2(b) shows the I(H) data acquired at V = +5mV. Over the ~350s needed to ramp H down from -2000 Oe to ~0 Oe, the MTJ remains in a P magnetic state, with $I_P$<0 despite V>0. In the MTJ's AP state, $I_{AP}$>0 over ~310s. The abrupt magnetic field dependence of the switch in sign of current clearly shows that the current sign change originates from the change in the MTJ's magnetic state, and not the magnetic field amplitude/sweep. Both $I_P$ and $I_{AP}$ exceed the maximum 500pA possible experimental offset by nearly 2 orders of magnitude (see SI). Thus, in this MTJ, the direction of static current flow can be reversed by simply switching the MTJ's spintronic state.

Since the external magnetic field is static, and we do not expect a spin texture in our FM electrodes, a spin motive force explanation[1,29,30] seems unlikely. In these and our experiments, no explicit temperature difference between electrodes, or temperature gradient, is applied to the device, such that a spin caloritronics[31] explanation, while possible, is not obvious. We further discuss in the Methods and SI how photovoltage/photocurrent and conventional/spintronic thermovoltage[32,33] artifacts can be excluded here.

These I(H) datapoints are confirmed through I(V) measurements at RT in the MTJ's P and AP states (see Fig. 2(c)), which reveal the following features: 1) at V = 0, $I_P \neq I_{AP} \neq 0$, with an amplitude that also exceeds any experimental offset by nearly 2 orders of magnitude; 2) a non-zero applied bias V leading to a measured I = 0 whose amplitude depends on the MTJ's magnetic state; 3) power generation above 0.1nW whose bias dependence depends on the MTJ's magnetic state, with a maximum current $I_{AP} \approx -70$nA at V=+1.4mV (see panel (c) inset); 4) bias-driven oscillations in current that depend on the MTJ's P/AP magnetic state, and thus on spin-dependent transport; 5) a bias range for which $I_P$ and $I_{AP}$ are of opposite sign, leading to TMR<-100%. These features of the $0 \rightarrow +V_{max} \rightarrow -V_{max} \rightarrow +V_{max} \rightarrow -V_{max} \rightarrow 0$ I(V) sweep are reproduced with high fidelity in Fig. 2c for another such sweep with differing maximum applied bias $V_{max}$ and bias step (i.e. a differing effective bias sweep rate), as well as by additional datasets (see SI). This eliminates any junction instability/memristive explanation[34]. 6) The numerically derived junction conductance dI/dV of the data of Fig. 2(c), shown in Fig. 2(d), reveals spintronically determined conductance jumps, and spectral features as low as 0.25meV --- despite the $2k_B T \approx 50$ meV thermal smearing expected at 295K --- that are statistically beyond the error bar (see Fig. S4 of the SI) thanks to an excellent signal-to-noise ratio. This spectral sharpness is also witnessed through a 300% TMR peak with a full-width-half-max of ~1.3meV (see Fig. 2(e)), which arises from a combination of local maxima(minima) in $I_P(I_{AP})$ at V = -3.5 mV. This non-optimized device's spintronic performance at 295K rivals the 600% record for FeCoB/MgO-class MTJs --- obtained through a 20-fold performance increase over 7 years[35,36] --- but since the Co electrodes cannot be *bcc*-oriented here[16,37] (see Methods), only a new mechanism can explain this spintronic performance.

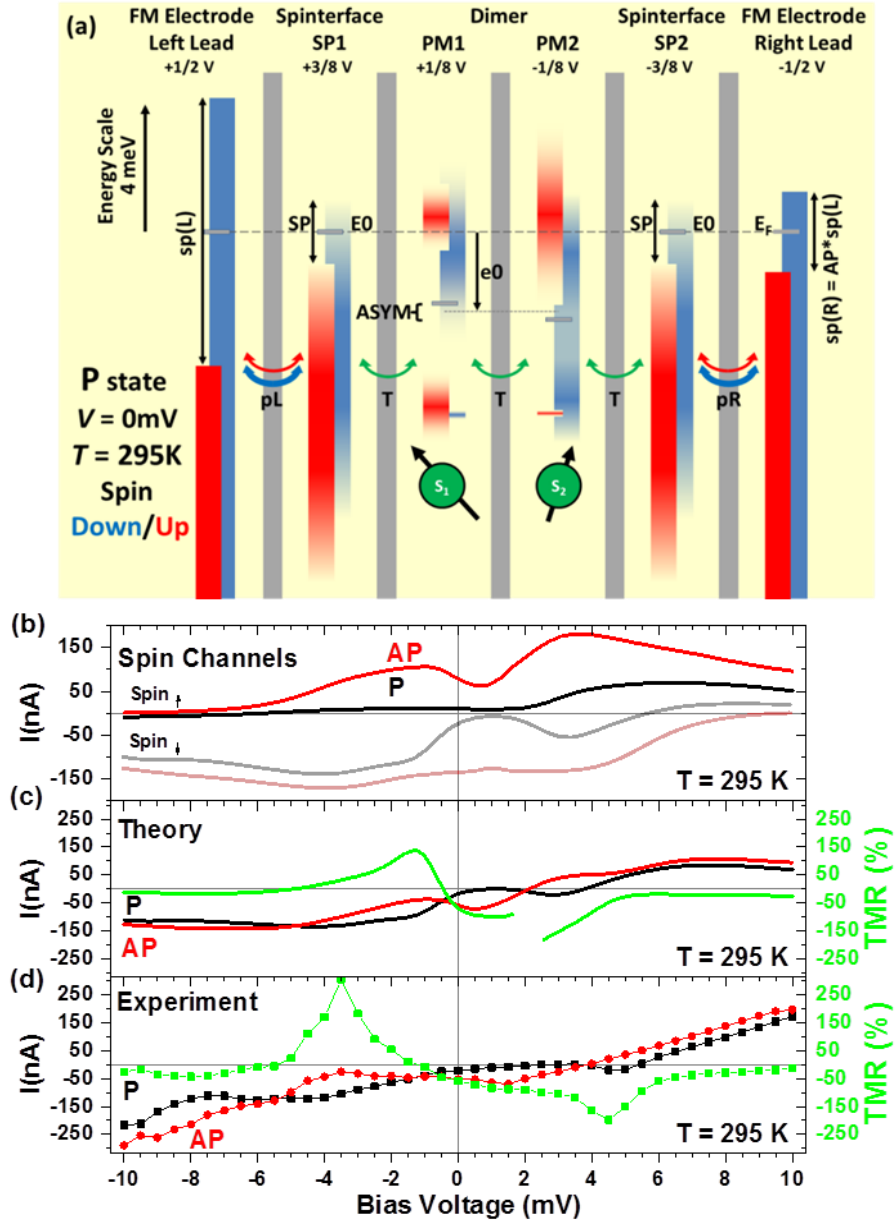

Figure 3: **Analytical theory linking experiments to the spin engine.** (a) Schematic of the analytical model of spin-conserved transport across a PM dimer (PM1 & PM2) separated from each FM lead by a spinterface. The calculated case of the P magnetic state at V=0 is shown. (b) Theoretical bias dependence of spin ↑ (solid) and spin ↓ (semi-transparent) current in the MTJ's P (black) and AP(red) magnetic states. (c) Theoretical and (d) experimental bias dependencies of current in the MTJ's P (black) and AP (red) magnetic states, and of the resulting TMR (green), using the same current/TMR scales. A same set of parameters (see Methods for details) was used to describe the MTJ's P/AP outer properties: *sp*=8.4, *SP*=2, *pL*=*pR*=0.35, *AP*=0.3, $E_0$=0, with pR and AP changing sign upon P↔AP, while we introduced minor variations in the PM dimer's starting conditions for P(AP): $e_0$=-2.5(+0.85) and *ASYM*=0.5(-0.75). Parameters are described in the Methods. Within an apparent factor in the voltage scale, the analytical model including the out-of-equilibrium hypothesis strongly mimicks the RT experiment, thereby linking it to the spin engine (Fig. 1).

This spectral sharpness in magnetotransport features at RT despite the expected thermal broadening, and the excellent signal-to-noise ratio, constitute a direct experimental signature of our spin engine at work. Indeed, only PM centers in the barrier, protected by the spinterfaces and tunnel barrier against an electrode-induced electronic broadening, could generate such spectrally sharp features with an energy position and amplitude that depend on the MTJ's P/AP magnetic state. From $k_B$T≈0.25 meV, we estimate an effective electronic temperature of the PM centers of 3K. This cooling is the manifestation of harvesting energy from the PM centers' spin fluctuations upon spin rectification in the junction.

Since a non-zero current is present at V=0 across this normally passive component, we observe that the MTJ is intrinsically out of equilibrium. Consequently, to further link our experimental results with our conceptual spin engine, we analytically consider an out-of-equilibrium nanotransport path across the MTJ comprising two PM centers (see Fig. 3(a) and the Methods/SI for details). Their initially discrete energy levels (gray lines) are broadened to form a PM dimer as bonding/anti-bonding and spin degeneracies are lifted (see PM 1&2 of Figs. 3(a)) through a magnetic exchange coupling that is bias-dependent[20,38]. To place the junction out of equilibrium, we impose a spin splitting of the FM electrodes' chemical potential. Following our spin engine proposal, our analytical model's magnetic interactions (with Heisenberg, Ising and Dzyaloshinskii–Moriya contributions) result in a spintronic anisotropy[19,20] onto PM1 and PM2 due to the spatially orientated, spin-polarized DOS of each FM electrode, as mediated by spinterfaces (SP1 & SP2).

Consistently with our experimental results, we constrain the model's 7-fold fitting parameter space using the following physical requirements: 1) the parameters should realistically describe the MTJ's outer properties (FM electrode + spinterface), including a higher spin polarization at the lower Co/MgO MTJ interface due to C dusting[24] (see Methods); 2) these outer properties should remain identical in the MTJ's P/AP magnetic states; 3) only minor changes to the PM dimer's properties are allowed between the P and AP cases. To account for T = 295K, the FM electrodes' Fermi level is broadened by 26meV (not shown in Fig. 3(a)). Fig. S7 shows the complex bias dependence of this spintronic potential landscape for each spinterface/PM center (SP1, PM1, PM2 & SP2), depending on the spin channel and the MTJ's P or AP magnetic state considered. Note how our model fulfills the spin engine's $\Delta < k_B T$ condition.

Referring to Fig 3(b), we observe a bias anti-symmetric imbalance in the oppositely propagating spin channels of current, which strongly depends on the MTJ's P/AP magnetic state. This leads to a sizeable spintronic difference in current, in particular at V=0. We recopy the $I_P(V)$, $I_{AP}(V)$ and TMR(V) experimental data of Fig. 2(c)/(e) as Fig. 3(d) in order to compare them with their analytical counterparts, shown in Fig. 3(c). Despite a skewed bias position that could underscore the simplicity of the bias voltage distribution (see Fig. 3(a)), our out-of-equilibrium analytical model reproduces all trends and salient features of the experimental magnetotransport data, including the spintronically dependent non-zero current at V=0, large TMR peak at V<0 and the bias region for V>0 with differing signs of $I_P$ and $I_{AP}$. A degraded agreement at large V likely reflects how our model only considers sequential transport across the 4 QDs, and not direct transport between the FM electrodes, which can become significant as the QD levels are energetically shifted away from one another. With a level of fidelity between experiment and theory that is unsurpassed for a tunneling spintronic device (compare with e.g. refs [17,18,39–41]), our analytical model strongly links our experimental results to our spin engine concept.

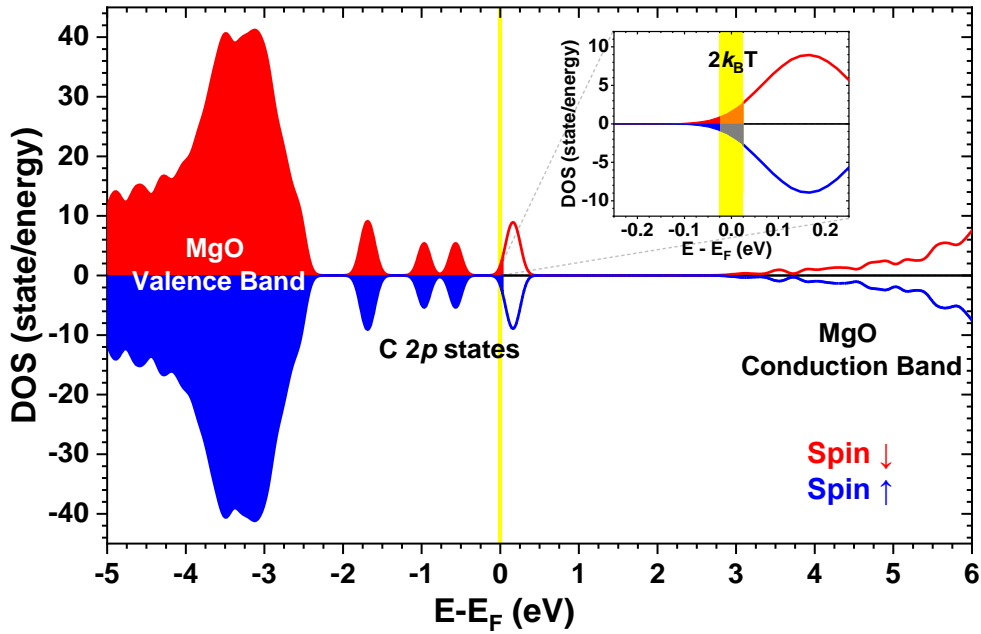

Figure 4: **Origin of the experiment's PM centers.** Ab-initio calculations of the spin-resolved DOS of MgO containing a carbon dimer in 4[th] nearest-neighbor configuration. The paramagnetic, AF-coupled C dimer generates energy levels around the MTJ's Fermi level $E_F$, including four spin-degenerate states that intersect $E_F$, thereby reproducing the analytical model's description of PM centers 1 & 2. See SI for the determination of the MTJ's $E_F$ energy position.

In general, compared to low-temperature transport across well-characterized quantum objects (e.g. from single atoms and dimers to molecules and atomic clusters[42–49]) thanks to a scanning tunnelling microscope (STM), it is thus far difficult to assemble and ascertain the effective nanotransport path[50] in a solid state device, especially for the oxides used as MTJ barriers. Here, uncontrolled imperfections such as oxygen vacancies in the MgO tunnel barrier can concentrate electronic tunnelling transport across a macrojunction onto a nanotransport path[50,51], such that the device operates due to a rare tunneling event[52,53]. This is what enables[54] the spin transfer torque effect underscoring key MTJ-based spintronic technologies[55]. While descriptions of the PM dimer in terms of Mn atoms or oxygen vacancies are much less likely here (see SI), paramagnetic C atoms occupying oxygen vacancy sites in MgO are possible considering our MTJ stack with C-dusted MgO interfaces. Indeed, carbon capture by single/double oxygen vacancies, which are present in our MgO[17,54], is energetically favorable (see SI and ref.[56]) and can yield both paramagnetic monomers (see SI) and dimers[28].

Our ab-initio theory shows that the C-C distance is crucial in order to reproduce our analytical model's results: only in a 4[th] nearest-neighbor positioning does the C dimer simultaneously exhibit AF coupling (favorable over FM by 0.125 eV, i.e. above experimental $k_B$T) and generate four states around the Fermi level $E_F$ of a Co/MgO/Co MTJ (see Fig. 4). On the other hand, C pairs in 1[st], 2[nd], 3[rd] and 5[th] nearest-neighbor configuration generate a FM state (see SI), which would be inconsistent with our analytical model. This stringent C impurity positional requirement on the oxygen sublattice might explain why our spin engine was experimentally observed only once out of ~200 attempts (see SI).

To increase reproducibility, we propose that all spintronic selector tracks be attempted (see introduction), noting that, in addition to the two published reports[1,2], similar effects were observed at low temperature on MTJs with manganite half metals[57]. In all cases, control over the spatial position and density of the barrier's PM centers will be required with a precision that, at this time, remains the domain of model STM-assembled junctions[45,46,58]. Considering that all reports involved microscale devices, this suggests reducing the junction's lateral size from the micro- to the nano-scale. In an oxide track, one may study tunnel barriers in which an oxygen vacancy-rich central region --- achieved e.g. by varying oxygen concentration in an Ar sputtering plasma during growth, is nominally seeded with impurities to be trapped by these vacancies as PM centers. Control over the electronic properties of, and magnetic interactions[59] between, PM centers in molecules suggests another, organic-based track using spintronic nanojunctions[60,61]. Whatever the route, the PM center(s) should experience dominant tunneling from one spintronic selector in order to adopt that selector's spin referential (see Fig. 1(b) and discussion). This can arise by tuning the selector/PM center tunneling rate through the insertion of an oxide/organic interlayer. Cleverly crafted operando techniques[50] that can directly characterize the PM center's properties within the device's nanotransport path can boost research efficiency. Overall, MgO spintronics represents a compelling route. Indeed, it benefits from both industrial penetration[62,63] and knowledge on how oxygen vacancies craft the spintronic nanotransport path[17,18,54], boasts lateral sizes down to 4.3nm[64], and has been conjugated with half-metallic electrodes operating at RT[12].  PM centers can be formed in MgO by trapping C, N or Si on oxygen vacancies (see Fig. 4, Ref. [28] and SI).

While our combined experiments and theory fully corroborate our spin engine proposal, we briefly note in the SI that the rectification of thermal fluctuations, which are obviously present experimentally and analytically, can generate work when combined with an increase in the entropy of quantum information[65] arising from spin transport onto the PM center due to its spin fluctuations. A quantum thermodynamical theory along a similar spintronic path has been proposed[8], while classical electronic implementations using capacitively coupled quantum dots have been demonstrated at low temperature[3,5,66]. Our results should thus generate research initiatives on quantum electronic circuits at the rather unexplored intersection between quantum thermodynamics[7,65,67–69] and spintronics[16,23,31,55,62]. More generally, our work also indicates that spinterfaces represent a compelling approach to integrating the quantum properties of nano-objects within a solid-state device's operation at room temperature, beyond proof-of-concept electronic decoupling strategies[58,70].  Indeed, the 300% TMR measured at RT, and our conceptual/theoretical framework, confirm the very high spin polarization of the organic spinterface at RT within magnetotransport across a solid-state device, i.e. validate its use as a spintronic selector.

The MTJ used to demonstrate our spin engine is an industrial-grade microelectronic device class used as the read head of hard disk drives, and with promising potential toward low-power information storage[62] and bio-inspired computing[55,71]. If routine reproducibility can be achieved, then for a typical STT-MRAM[63] 2D array of 50nm-wide MTJs with a pitch of 90nm, for 0.1nW per MTJ, and even assuming 1% efficiency due to engineering issues, the resulting power density would still exceed the raw solar power density on the earth surface by a factor of six. As an example, a suitable series/parallel assembly on a 1cm$^2$ chip could deliver ~58.8W of power to a 0.1$\Omega$ load at 3.5V. See SI for more details. Industrializing the MTJ's spin engine functionality through further research could fundamentally alter our energy-driven society's global constructs and climate change mitigation strategies.  Indeed, it is more versatile than solar cells: by sipping energy from the magnetic state within a calibrated thermal envelope, this spin engine constitutes an autonomous, essentially limitless, always-on source of electrical power without the constraints of solar panel deployment, exposure and grid balancing. It can be assembled using common-place materials, thereby easing global tensions over mineral extraction. Successful up-scaling could displace many implementations of chemical energy storage and their associated environmental impact, thereby reducing e-waste. This spin engine has the potential to alter the global economic, social and geopolitical world order through unfettered access to energy.

## Methods

Ta(5)/Co(10)/IrMn(7.5)/Co(4)/C($d_1$=0, 0.3, 0.6, 0.9)/MgO(2.5nm)/C($d_2$=0, 0.3, 0.6)/Co(10)/Pt(4)   samples (all thicknesses in nm) were sputter-grown on Corning 1737 glass substrates[72]. Stacks were post-annealed in an in-plane magnetic field of 200 Oe for 1 h at a temperature Ta of 200 °C to magnetically pin the lower electrode thanks to the IrMn antiferromagnetic layer. This low annealing temperature precludes the diffusion of Mn into the barrier[73], though it can promote C diffusion into MgO[56]. Samples were then processed by optical lithography[74], and measured on a variable-temperature magnetotransport bench. Only 1 out of 168 20 μm-diameter MTJs tested at T = 295K, with $d_1$=0.9 and $d_2$=0, a ten-fold larger RA product and anomalous TMR<0 at V=+10mV, revealed these peculiar power generation features and high TMR. An additional 48 MTJs did not exhibit I≠0 at V=0. We presume that annealing-induced C migration into MgO generated the MTJ's spinterfaces and PM centers.

Measurements on this MTJ were conducted over 90 minutes in 4-point measurement mode within a dark cryostat that remained between 295.3K and 294.5K with sample heater off. Furthermore, these peculiar transport features strongly depend on the MTJ's P/AP magnetic state. We can therefore exclude thermovoltage photovoltage/photocurrent, as well as any conventional or spintronic thermovoltage explanations. We discuss these discarded artifact sources in more detail in the SI.

The analytical model symmetrically segments the MTJ's bias drop into 6 zones around the junction mid-point, for which V≡0, using the same bias sign convention as in experiment. Current flow across the left-hand(right-hand) FM electrode is modeled by a spin-splitting $sp$($sp$ x $AP$) of its chemical potential. The DOS of the spinterfaces SP1 & SP2 consists in 10meV-wide bands that are centered around $E_F$ at V = 0 and are spin-split by $SP$. A constant E0=0 was used. PM 1&2 model the paramagnetic dimer as two spin states $S_1$ and $S_2$ that, *initially*, are energetically discrete, are positioned *eo* away from EF at V=0, and are energy-split by *ASYM* but are not spin-split. Current flows between the FM electrodes across SP1/PM1/PM2/SP2 through a tunneling rate T, which was fixed at [1 1 1] between SP1/PM1 ; PM1/PM2 ; PM2/SP2. Finally, pL(pR) describes a possible spin polarization of the tunelling transmission between the left(right) FM lead and SP1(SP2). The MTJ's AP state is described by switching the sign of pR and AP, i.e. by flipping the right-hand FM electrode magnetization. This experimentally corresponds to the free Co layer of the top FM electrode. AP=0.3 is consistent with an experimentally larger spin polarization of the C-dusted Co lower FM electrode[24], and to $d_1 \neq d_2$. The SI further details the model's transport formalism.

Within density functional theory, the electronic properties of the C dimer within MgO were computed using 64-atom supercells with a simple cubic structure with two substitutional carbon atoms in various configurations (see Fig.S7). These calculations were done using VASP code[75] based on the projector augmented wave (PAW) method[76] and the Pedrew, Burke, Enzerhof (PBE)[77] generalized gradient approximation for the exchange-correlation potential. The kinetic energy cutoff value of 500 eV for the plane wave basis set and the convergence criterion for the total energy of $10^{-8}$ eV is used. The carbon-doped structures are fully relaxed using a conjugate-gradient algorithm, such that the forces acting on atoms be less than 0.001 eV/Å. A $k$-point mesh of 6x6x6 with the Methfessel-Paxton method with a smearing τ= 0.1 eV is used. See SI for the determination of EF within a Co/MgO(12ML, *i.e.* ~2.5nm)/Co MTJ.

## Acknowledgements

We are grateful to J. Koski, G. Weick, K. McKenna and J. Blumberger for useful discussions, and to Y. Henry for carefully reading our manuscript. Devices were synthesized at the STNano technological platform. We acknowledge financial support from the Institut Carnot MICA (project 'Spinterface'), from the ANR (ANR-09-JCJC-0137, ANR-14-CE26-0009-01), the Labex NIE "Symmix" (ANR-11-LABX-0058 NIE) and Vetenskapsrådet). This work was performed using HPC resources from the Strasbourg Mesocenter and from the GENCI-CINES Grant 2016-gem1100.

## Author Contributions

M.B., E.B. and S.B. conceived the initial experiment. M.H. grew the sample stack. E.U., F.S., K.K. and J.A. implemented technological processing to make MTJs. E.U., F.S. and K.K. performed magnetotransport measurements. M.B., K.K., S.B., D.L., E.B., M.A. and W.W. analyzed magnetotransport results. B.V. and D.S. performed auxiliary measurements. J.F. designed the analytical models with input from M.B. K.K. applied the analytical models, with input from M.B. and J.F. B.T. and M.A. implemented DFT calculations with input from M.B. M.B., K.K. and B.T. prepared the manuscript. All authors commented on the manuscript.

## Competing Interests

The authors declare no competing interests.

## References

1. Hai, P. N., Ohya, S., Tanaka, M., Barnes, S. E. & Maekawa, S. Electromotive force and huge magnetoresistance in magnetic tunnel junctions. *Nature* **458**, 489 (2009).

2. Miao, G.-X., Chang, J., Assaf, B. A., Heiman, D. & Moodera, J. S. Spin regulation in composite spin-filter barrier devices. *Nat. Commun.* **5**, 3682 (2014).

3. Thierschmann, H. *et al.* Three-terminal energy harvester with coupled quantum dots. *Nat. Nanotechnol.* **10**, 854–858 (2015).

4. Jaliel, G. *et al.* Experimental realization of a quantum dot energy harvester. *ArXiv190110561 Cond-Mat* (2019).

5. Koski, J. V., Kutvonen, A., Khaymovich, I. M., Ala-Nissila, T. & Pekola, J. P. On-Chip Maxwell's Demon as an Information-Powered Refrigerator. *Phys. Rev. Lett.* **115**, 260602 (2015).

6. Sánchez, R. & Büttiker, M. Optimal energy quanta to current conversion. *Phys. Rev. B* **83**, 085428 (2011).

7. Strasberg, P., Schaller, G., Brandes, T. & Esposito, M. Quantum and Information Thermodynamics: A Unifying Framework Based on Repeated Interactions. *Phys. Rev. X* **7**, 021003 (2017).

8. Ptaszyński, K. Autonomous quantum Maxwell's demon based on two exchange-coupled quantum dots. *Phys. Rev. E* **97**, 012116 (2018).

9. Wang, W.-B. *et al.* Realization of Quantum Maxwell's Demon with Solid-State Spins *. *Chin. Phys. Lett.* **35**, 040301 (2018).

10. Bowen, M. *et al.* Half-metallicity proven using fully spin-polarized tunnelling. *J. Phys. Condens. Matter* **17**, L407–L409 (2005).

11. Liu, H. *et al.* Giant tunneling magnetoresistance in epitaxial Co2MnSi/MgO/Co2MnSi magnetic tunnel junctions by half-metallicity of Co2MnSi and coherent tunneling. *Appl. Phys. Lett.* **101**, 132418 (2012).

12. Boehnke, A. *et al.* Large magneto-Seebeck effect in magnetic tunnel junctions with half-metallic Heusler electrodes. *Nat. Commun.* **8**, 1626 (2017).

13. Ashton, M. *et al.* Two-Dimensional Intrinsic Half-Metals With Large Spin Gaps. *Nano Lett.* **17**, 5251–5257 (2017).

14. Miao, G.-X. & Moodera, J. S. Spin manipulation with magnetic semiconductor barriers. *Phys. Chem. Chem. Phys.* **17**, 751–761 (2015).

15. Delprat, S. *et al.* Molecular spintronics: the role of spin-dependent hybridization. *J. Phys. Appl. Phys.* **51**, 473001 (2018).

16. Miao, G.-X., Münzenberg, M. & Moodera, J. S. Tunneling path toward spintronics. *Rep. Prog. Phys.* **74**, 036501 (2011).

17. Schleicher, F. *et al.* Localized states in advanced dielectrics from the vantage of spin- and symmetry-polarized tunnelling across MgO. *Nat. Commun.* **5**, 4547 (2014).

18. Taudul, B. *et al.* Tunneling Spintronics across MgO Driven by Double Oxygen Vacancies. *Adv. Electron. Mater.* 1600390 (2017). doi:10.1002/aelm.201600390

19. Misiorny, M., Hell, M. & Wegewijs, M. R. Spintronic magnetic anisotropy. *Nat. Phys.* **9**, 801–805 (2013).

20. Fransson, J., Ren, J. & Zhu, J.-X. Electrical and Thermal Control of Magnetic Exchange Interactions. *Phys. Rev. Lett.* **113**, 257201 (2014).

21. Velev, J., Dowben, P., Tsymbal, E., Jenkins, S. & Caruso, A. Interface effects in spin-polarized metal/insulator layered structures. *Surf. Sci. Rep.* **63**, 400–425 (2008).

22. Barraud, C. *et al.* Unravelling the role of the interface for spin injection into organic semiconductors. *Nat. Phys.* **6**, 615–620 (2010).

23. Raman, K. V. Interface-assisted molecular spintronics. *Appl. Phys. Rev.* **1**, 031101 (2014).

24. Djeghloul, F. *et al.* Highly spin-polarized carbon-based spinterfaces. *Carbon* **87**, 269–274 (2015).

25. Djeghloul, F. *et al.* High Spin Polarization at Ferromagnetic Metal-Organic Interfaces: a Generic Property. *J. Phys. Chem. Lett.* **7**, 2310–2315 (2016).

26. Cinchetti, M., Dediu, V. A. & Hueso, L. E. Activating the molecular spinterface. *Nat. Mater.* **16**, 507–515 (2017).

27. Valet, T. & Fert, A. Theory of the perpendicular magnetoresistance in magnetic multilayers. *Phys. Rev. B* **48**, 7099 (1993).

28. Wu, H. *et al.* Magnetism in C- or N-doped MgO and ZnO: A Density-Functional Study of Impurity Pairs. *Phys. Rev. Lett.* **105**, 267203 (2010).

29. Barnes, S. E. & Maekawa, S. Generalization of Faraday's Law to Include Nonconservative Spin Forces. *Phys. Rev. Lett.* **98**, 246601 (2007).

30. Tanabe, K. *et al.* Spin-motive force due to a gyrating magnetic vortex. *Nat. Commun.* **3**, 845 (2012).

31. Hoffmann, A. & Bader, S. D. Opportunities at the Frontiers of Spintronics. *Phys. Rev. Appl.* **4**, 047001 (2015).

32. Walter, M. *et al.* Seebeck effect in magnetic tunnel junctions. *Nat. Mater.* **10**, 742–746 (2011).

33. Jaramillo, J. D. V. & Fransson, J. Charge Transport and Entropy Production Rate in Magnetically Active Molecular Dimer. *J. Phys. Chem. C* **121**, 27357–27368 (2017).

34. Baeumer, C. *et al.* Quantifying redox-induced Schottky barrier variations in memristive devices via *in operando* spectromicroscopy with graphene electrodes. *Nat. Commun.* **7**, 12398 (2016).

35. Bowen, M. *et al.* Large magnetoresistance in Fe/MgO/FeCo(001) epitaxial tunnel junctions on GaAs(001). *Appl. Phys. Lett.* **79**, 1655 (2001).

36. Ikeda, S. *et al.* Tunnel magnetoresistance of 604% at 300K by suppression of Ta diffusion in CoFeB∕MgO∕CoFeB pseudo-spin-valves annealed at high temperature. *Appl. Phys. Lett.* **93**, 082508 (2008).

37. Yuasa, S., Fukushima, A., Kubota, H., Suzuki, Y. & Ando, K. Giant tunneling magnetoresistance up to 410% at room temperature in fully epitaxial Co∕MgO∕Co magnetic tunnel junctions with bcc Co(001) electrodes. *Appl. Phys. Lett.* **89**, 042505 (2006).

38. Saygun, T., Bylin, J., Hammar, H. & Fransson, J. Voltage-Induced Switching Dynamics of a Coupled Spin Pair in a Molecular Junction. *Nano Lett.* **16**, 2824–2829 (2016).

39. Bowen, M. *et al.* Observation of Fowler–Nordheim hole tunneling across an electron tunnel junction due to total symmetry filtering. *Phys. Rev. B* **73**, 140408(R) (2006).

40. Matsumoto, R. *et al.* Spin-dependent tunneling in epitaxial Fe/Cr/MgO/Fe magnetic tunnel junctions with an ultrathin Cr(001) spacer layer. *Phys. Rev. B* **79**, 174436 (2009).

41. Teixeira, J. M. *et al.* Resonant Tunneling through Electronic Trapping States in Thin MgO Magnetic Junctions. *Phys. Rev. Lett.* **106**, 196601 (2011).

42. Natterer, F. D. *et al.* Reading and writing single-atom magnets. *Nature* **543**, 226–228 (2017).

43. Atzori, M. *et al.* Room-Temperature Quantum Coherence and Rabi Oscillations in Vanadyl Phthalocyanine: Toward Multifunctional Molecular Spin Qubits. *J. Am. Chem. Soc.* **138**, 2154–2157 (2016).

44. Muenks, M., Jacobson, P., Ternes, M. & Kern, K. Correlation-driven transport asymmetries through coupled spins in a tunnel junction. *Nat. Commun.* **8**, 14119 (2017).

45. Ormaza, M. *et al.* Efficient Spin-Flip Excitation of a Nickelocene Molecule. *Nano Lett.* **17**, 1877–1882 (2017).

46. Loth, S. *et al.* Controlling the state of quantum spins with electric currents. *Nat. Phys.* **6**, 340–344 (2010).

47. Hermenau, J. *et al.* A gateway towards non-collinear spin processing using three-atom magnets with strong substrate coupling. *Nat. Commun.* **8**, 642 (2017).

48. Moreno-Pineda, E., Godfrin, C., Balestro, F., Wernsdorfer, W. & Ruben, M. Molecular spin qudits for quantum algorithms. *Chem. Soc. Rev.* **47**, 501–513 (2018).

49. Casola, F., van der Sar, T. & Yacoby, A. Probing condensed matter physics with magnetometry based on nitrogen-vacancy centres in diamond. *Nat. Rev. Mater.* **3**, 17088 (2018).

50. Studniarek, M. *et al.* Probing a Device's Active Atoms. *Adv. Mater.* 1606578 (2017). doi:10.1002/adma.201606578

51. Kim, D. J. *et al.* Control of defect-mediated tunneling barrier heights in ultrathin MgO films. *Appl. Phys. Lett.* **97**, 263502 (2010).

52. Bardou, F. Rare events in quantum tunneling. *Europhys. Lett. EPL* **39**, 239–244 (1997).

53. Da Costa, V., Tiusan, C., Dimopoulos, T. & Ounadjela, K. Tunneling Phenomena as a Probe to Investigate Atomic Scale Fluctuations in Metal/Oxide/Metal Magnetic Tunnel Junctions. *Phys. Rev. Lett.* **85**, 876–879 (2000).

54. Halisdemir, U. *et al.* Oxygen-vacancy driven tunnelling spintronics across MgO. in *SPIE* (eds. Drouhin, H.-J., Wegrowe, J.-E. & Razeghi, M.) **9931**, 99310H (2016).

55. Locatelli, N., Cros, V. & Grollier, J. Spin-torque building blocks. *Nat. Mater.* **13**, 11–20 (2013).

56. Tiusan, C. *et al.* Spin tunnelling phenomena in single-crystal magnetic tunnel junction systems. *J. Phys. Condens. Matter* **19**, 165201 (2007).

57. Bowen, M. Experimental Insights into Spin-Polarized Solid State Tunneling. (Universite de Paris XI, 2003).

58. Paul, W. *et al.* Control of the millisecond spin lifetime of an electrically probed atom. *Nat. Phys.* **13**, 403–407 (2017).

59. Serri, M. *et al.* High-temperature antiferromagnetism in molecular semiconductor thin films and nanostructures. *Nat. Commun.* **5**, 3079 (2014).

60. Barraud, C. *et al.* Unidirectional Spin-Dependent Molecule-Ferromagnet Hybridized States Anisotropy in Cobalt Phthalocyanine Based Magnetic Tunnel Junctions. *Phys. Rev. Lett.* **114**, 206603 (2015).

61. Barraud, C. *et al.* Phthalocyanine based molecular spintronic devices. *Dalton Trans* **45**, 16694–16699 (2016).

62. Kent, A. D. & Worledge, D. C. A new spin on magnetic memories. *Nat. Nanotechnol.* **10**, 187–191 (2015).

63. Chung, S.-W. *et al.* 4Gbit density STT-MRAM using perpendicular MTJ realized with compact cell structure. in *2016 IEEE International Electron Devices Meeting (IEDM)* 27.1.1-27.1.4 (IEEE, 2016). doi:10.1109/IEDM.2016.7838490

64. Watanabe, K., Jinnai, B., Fukami, S., Sato, H. & Ohno, H. Shape anisotropy revisited in single-digit nanometer magnetic tunnel junctions. *Nat. Commun.* **9**, (2018).

65. Mandal, D. & Jarzynski, C. Work and information processing in a solvable model of Maxwell's demon. *Proc. Natl. Acad. Sci.* **109**, 11641–11645 (2012).

66. Josefsson, M. *et al.* A quantum-dot heat engine operated close to thermodynamic efficiency limits. *ArXiv171000742 Cond-Mat* (2017).

67. Parrondo, J. M. R., Horowitz, J. M. & Sagawa, T. Thermodynamics of information. *Nat. Phys.* **11**, 131–139 (2015).

68. Hänggi, P. & Talkner, P. Fluctuation theorems go beyond the linear response regime to describe systems far from equilibrium. But what happens to these theorems when we enter the quantum realm? The answers, it seems, are now coming thick and fast. *Nat. Phys.* **11**, 3 (2015).

69. Rio, L. del, Åberg, J., Renner, R., Dahlsten, O. & Vedral, V. The thermodynamic meaning of negative entropy. *Nature* **474**, 61–63 (2011).

70. Heinrich, B. W., Braun, L., Pascual, J. I. & Franke, K. J. Protection of excited spin states by a superconducting energy gap. *Nat. Phys.* **9**, 765–768 (2013).

71. Krzysteczko, P., Münchenberger, J., Schäfers, M., Reiss, G. & Thomas, A. The Memristive Magnetic Tunnel Junction as a Nanoscopic Synapse-Neuron System. *Adv. Mater.* **24**, 762–766 (2012).

72. Bernos, J. *et al.* Impact of electron-electron interactions induced by disorder at interfaces on spin-dependent tunneling in Co-Fe-B/MgO/Co-Fe-B magnetic tunnel junctions. *Phys. Rev. B* **82**, 060405(R) (2010).

73. Hayakawa, J., Ikeda, S., Lee, Y. M., Matsukura, F. & Ohno, H. Effect of high annealing temperature on giant tunnel magnetoresistance ratio of CoFeB∕MgO∕CoFeB magnetic tunnel junctions. *Appl. Phys. Lett.* **89**, 232510 (2006).

74. Halley, D. *et al.* Electrical switching in Fe∕Cr∕MgO∕Fe magnetic tunnel junctions. *Appl. Phys. Lett.* **92**, 212115 (2008).

75. Kresse, G. & Furthmüller, J. Efficient iterative schemes for *ab initio* total-energy calculations using a plane-wave basis set. *Phys. Rev. B* **54**, 11169–11186 (1996).

76. Kresse, G. & Joubert, D. From ultrasoft pseudopotentials to the projector augmented-wave method. *Phys. Rev. B* **59**, 1758–1775 (1999).

77.  Perdew, J. P., Burke, K. & Ernzerhof, M. Generalized Gradient Approximation Made Simple. *Phys. Rev. Lett.* **77**,

3865–3868 (1996).

# Spin-driven electrical power generation at room temperature

## *Supplementary Information*

# 1 Junction Statistics

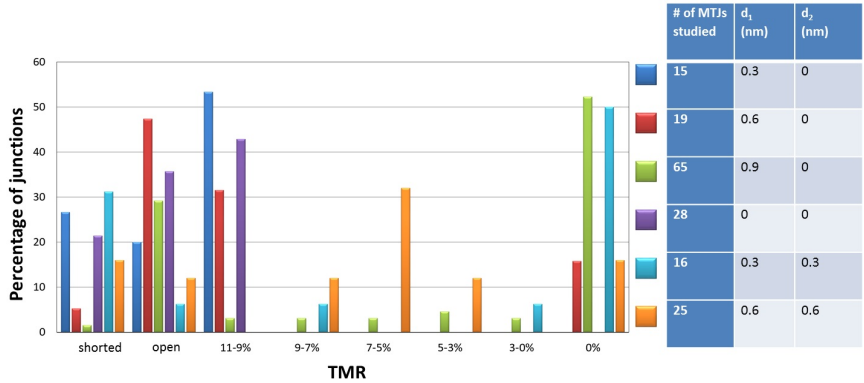

Fig.S 1: Transport statistics across C-dusted Co/MgO/Co MTJ stacks according to the nominal thickness $d_1$ and $d_2$ of C at the MTJ's bottom and top interfaces, respectively.

We studied 168 MTJs, grown by sputter deposition, with the nominal Glass // Ta(5) / Co(10) / IrMn(7.5 )/ Co(4) / C(d1) / MgO(2.5nm) / C(d2) / Co(10) / Pt(4) composition. Fig.S 1 shows an overview of the transport / magnetotransport properties of these 168 MTJs, and indicates the nominal thicknesses d1 and d2 of the C dusting layers at the MTJ stacks's bottom and top interfaces, respectively. Of the 168 MTJs, 102 were neither short-circuited nor open circuited. Aside from the MTJ described in the main text, this subset of MTJs exhibited standard transport characteristics, such as nearly linear IVs and low positive TMR at V= 10mV and T = 295K. Within an additional measurement campaign, we probed 48 additional MTJs, but found no anomalous non-zero current at V=0.

# 2 Additional datasets

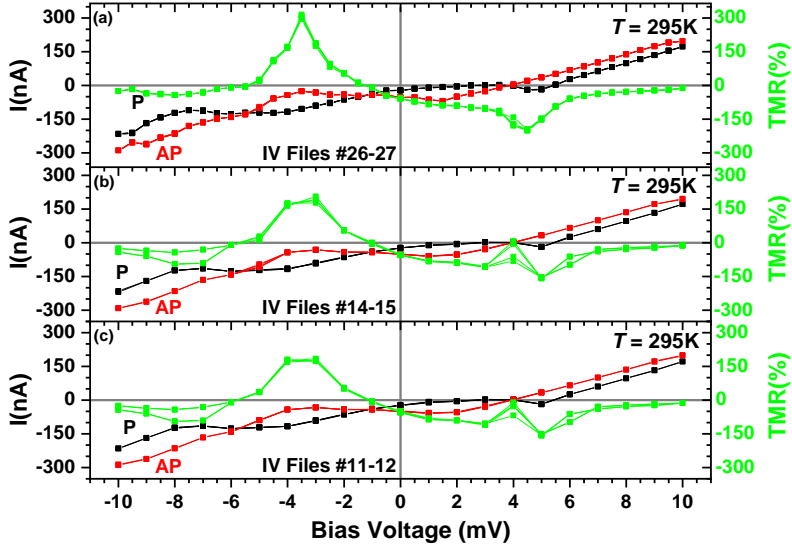

Fig.S 2: **Additional magnetotransport datasets at T = 295K up to ±10mV.** IV files 26-27, which are shown in Fig.2 of the main manuscript as closed symbols, were acquired with the Keithley 2636 in manual current range, while current autorange was on for IV files 11-12 and 14-15, for which the maximum V was 100mV and 30mV, respectively. For this MTJ, the autorange parameter caused a jump to a new IV branches at higher bias. To compare with panel (a), we therefore show in panels (b-c) only data from the junction state corresponding to that of IV files #26-17. The same current and TMR scale is used in all three panels. $0 \rightarrow +10\text{mV} \rightarrow -10\text{mV} \rightarrow +10\text{mV} \rightarrow -10\text{mV} \rightarrow 0$ sweeps are represented for every dataset (i.e. there are in fact 4 datapoints for every bias value).

Although the experimental data of the main manuscript was acquired on only one MTJ, its unusual magnetotransport features are highly reproducible. We begin by reproducing in Fig.S 2(a) the experimental IV data of Fig.2(c) and Fig.3(d) of the main manuscript. Panels (b-c) display the IV and TMR bias dependencies for IV files 14-15 and 11-12. Despite differences in bias step and maximum applied bias (see Figure caption for details), the magnetotransport features are quantitatively similar.

The magnetotransport features of Fig.S 2 are quantitatively reproduced through additional IV datasets shown in Fig.S 3. In each of the MTJ's P and AP magnetic state, we thus observe through seven separate $0 \rightarrow + \text{Vmax} \rightarrow -\text{Vmax} \rightarrow +\text{Vmax} \rightarrow -\text{Vmax} \rightarrow 0$ (i.e. for a total of 32 datapoints at a given V and MTJ magnetic state), the same unusual magnetotransport features. Note the high fidelity of each dataset despite the presence of multiple bias sweep in each panel.

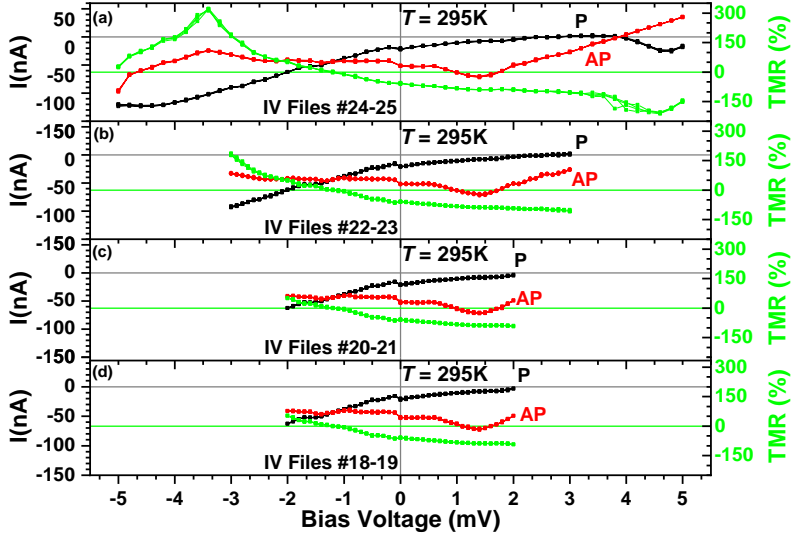

Fig.S 3: **Additional magnetotransport datasets at T = 295K up to ±5mV.** $0 \to +$ Vmax $\to -$Vmax $\to +$Vmax $\to -$Vmax $\to 0$ sweeps are represented for every dataset (i.e. there are in fact 4 datapoints for every bias value). All IV files (18-25) were acquired in manual current range. The data of panel (b) appears in Fig. 2 of the main manuscript as open symbols.

## 3 Controlling experimental artifacts

The MTJ was measured on the cold finger of an optically sealed cryostat. Since the sample remained in the dark during measurements, this ensures that no spurious photovoltage/photocurrent effects are taking place here. Measurements were performed using a Keithley 2636 Sourcemeter, in 4-point mode with the top electrode labelled as V+/I+. This latter point in principle eliminates possible thermovoltage artifacts along the leads. Separate testing using calibrated resistances revealed a maximum current offset of 500pA and a maximum voltage offset of 0.1 mV. We observe that the current across the MTJ can reach ≈ -50nA at V=0 (AP magnetic state), while several bias voltage values leading to I=0 are observed in the MTJ's P magnetic state. No shifting in either V and/or I (presumably to correct for a V or I offset) can prevent the dataset from entering a 'forbidden' IV quadrant (with I and V of opposite sign).

Furthermore, these anomalies strongly depend on the MTJ's magnetic state. When sweeping the external magnetic field H, the current remains constant as long as the MTJ's magnetic state is unchanged (see Fig. 2(a-b)). However, when changing H causes the MTJ's state to switch (P ⟷ AP), a change in current is observed. While this is normal for a MTJ (i.e. due to spintronic performance), note how the non-zero current at V=0 also changes. Note also how, at V=+5mV, switching the MTJ's magnetic state can change the sign of measured current (see Fig. 2(b)). This dependence of the non-zero current at V=0, or of the sign of current at constant V, upon the MTJ's magnetic state further discredits an explanation in terms of thermovoltage along the leads.

To avoid any temperature gradient effect, the experiment was conducted at RT and the sample heater was turned off for several hours prior to measurement. The dataset's 3 I(H)'s and 17 I(V)'s (files #10-29) were acquired over 90 minutes. Within that time interval, the temperature measured ≈15cm away from the MTJ on the cryostat's Cu cold finger decreased monotonously from 295.3K to 294.46K, i.e. at a rate of ≈0.01K/min. Thus, the two P/AP $0 \to +10$mV $\to -10$mV $\to +10$mV $\to -10$mV $\to 0$ I(V) (file #26-27) of Fig. 2(c) were acquired sequentially with a starting temperature of 294.61K and 294.6K, which decreased by 0.01K over the 93s required by each measurement. The I(H) measurements of Fig. 2(a) (file #10, i.e. the first measurement reported here; T:295.3K→295.13K) and Fig. 2(b) (file #29, i.e. the last measurement reported here; T:294.53K →294.46K) initiate and conclude the reported dataset. In particular, Fig. 2(c) shows a quantitative agreement between the I(H) data of Fig. 2(a) (file #10; blue crosses in Fig. 2(c) ) and the I(V) data of Fig. 2(c) (file#26-27). This shows that the minor drift in temperature has no impact on the spintronic response reported here.

We have carefully considered possible memristance/ageing effects here. First the 0 → +Vmax → -Vmax → +Vmax → -Vmax → 0 IV measurement protocol can reveal whether the junction's state is evolving due to its electrical history. As seen in Figs.S2-S3, the data are highly reproducible if our Keithley 2636 remains within a fixed current range. We do not witness any opening of the IV loop associated with a memristance effect. We also observe quantitative agreement between IV files 11-12 and 26-27, despite having applied up to 100mV and performed multiple bias sweept between these initial and final datasets (see Fig.S 2). We therefore do not witness any device ageing within this dataset.

# 4 Effective 0.25meV experimental spectroscopic resolution at 295K despite 26meV expected thermal smearing

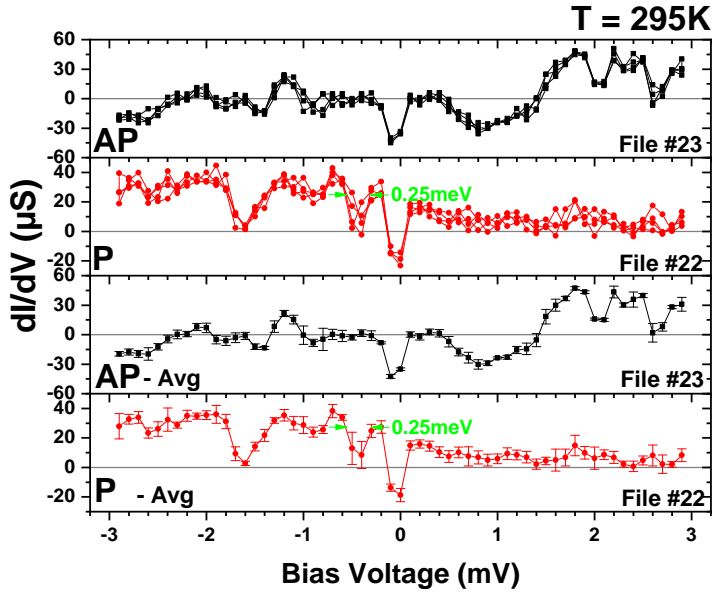

Fig.S 4: **Spectral features in bias dependent magnetotransport at 295K with sub-meV sharpness.** Numerical conductances dI/dV in the MTJ's (a) P (I(V) file #22) and (b) AP (I(V) file #23) states. Error bars are calculated as the standard deviation from the mean dI/dV. The four points at a given bias V, acquired through the 0 → +Vmax → -Vmax → +Vmax → -Vmax → 0 IV sweep, were used to derive these statistics. Features beyond the error bars as low as 0.25meV are observed. As needed, a mathematically erroneous point at V=0 was removed.

Referring to the dI/dV(V) graphs of Fig.2d, despite the 295K acquisition temperature and the absence of a lock-in technique, which resulted in a numerical derivation of dI/dV from I(V) data, we witness spectroscopic features with an energy resolution well below 26meV expected from thermal smearing of the MTJ electrodes' Fermi level. As shown in Fig.S 4, the spectral width of these features is as low as 0.25meV and is well beyond the errors bars.

We present in Fig.S5 a series of dI/dV data calculated from the raw IV data of files #18-23. Comparing these data, one concludes that, aside from very minor variations, spectrally sharp features are reproducible for a given MTJ magnetic state, and depend on the MTJ's magnetic state, i.e. are of spintronic origin.

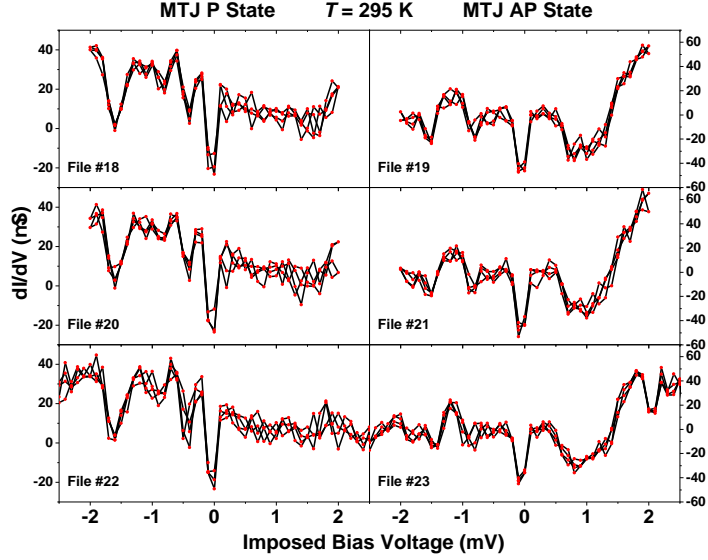

Fig.S 5: **Reproducibility of spectrally sharp spintronic features.** Numerical conductances dI/dV in the MTJ's P and AP states using I(V) files #18-23, each with a $0 \rightarrow +\mathrm{Vmax} \rightarrow -\mathrm{Vmax} \rightarrow +\mathrm{Vmax} \rightarrow -\mathrm{Vmax} \rightarrow 0$ sweep. As needed, a mathematically erroneous point at V=0 was removed.

## 5 Transport formalism

To analytically reproduce the spin engine's conceptual ingredients, we consider transport across two paramagnetic centers separated from ferromagnetic electrodes by spinterfaces, as schematized in Fig. S6. For simplicity, we refer to all four elements as quantum dots.

**Spin model** - We assume ferromagnetic leads modelled by $\mathcal{H}_\chi = \sum_{\mathbf{k}\sigma \in \chi} \varepsilon_{\mathbf{k}\sigma} c^\dagger_{\mathbf{k}\sigma} c_{\mathbf{k}\sigma}$, where $c^\dagger_{\mathbf{k}\sigma}$ ($c_{\mathbf{k}\sigma}$) creates (annihilates) an electron at the energy $\varepsilon_{\mathbf{k}\sigma}$ with momentum $\mathbf{k}$ and spin $\sigma = \uparrow, \downarrow$, and we reserve the notation $\mathbf{k} = \mathbf{p}$ ($\mathbf{k} = \mathbf{q}$) for electrons in the left, $\chi = L$, (right, $\chi = R$) lead. In the tunneling junction between the leads is a molecular assembly embedded comprising $M$ discrete levels coupled in series, here labelled $m = 1, \ldots, M$. To each of the levels there a localized spin $\mathbf{S}_m$ is coupled through exchange interaction, with rate $v_m$. Assuming a nearest neighbor interaction between the levels, we model the molecular assembly by $\mathcal{H}_M = \sum_{m\sigma} \varepsilon_{m\sigma} d^\dagger_{m\sigma} d_{m\sigma} + \sum_{m\sigma} (T_m d^\dagger_{m\sigma} d_{m+1\sigma} + H.c.) + \sum_m v_m \mathbf{s}_m \cdot \mathbf{S}_m$. Finally, the tunneling between the leads and the molecular assembly is modelled by $\mathcal{H}_T = \sum_{\mathbf{p}\sigma} T_{L\sigma} c^\dagger_{\mathbf{p}\sigma} d_{1\sigma} + \sum_{\mathbf{q}\sigma} T_{R\sigma} c^\dagger_{\mathbf{q}\sigma} d_{M\sigma} + H.c.$

**Charge transport** - Before we derive the expression for the charge current, we consider the models for the leads, captured by $\mathcal{H}_\chi = \sum_{\mathbf{k}\sigma} \varepsilon_{\mathbf{k}\sigma} n_{\mathbf{k}\sigma}$, where $n_{\mathbf{k}\sigma} = c^\dagger_{\mathbf{k}\sigma} c_{\mathbf{k}\sigma}$ is the number operator. As an open system, we have to work in terms of the grand canonical ensemble which implies the introduction of the chemical potential $\mu_\chi$ through the replacement $\sum_{\mathbf{k}\sigma} \varepsilon_{\mathbf{k}\sigma} n_{\mathbf{k}\sigma} \rightarrow \sum_{\mathbf{k}\sigma} (\varepsilon_{\mathbf{k}\sigma} - \mu_L) n_{\mathbf{k}\sigma}$. Furthermore, since the energy dispersion $\varepsilon_{\mathbf{k}\sigma}$ in the ferromagnetic leads can be expressed as $\varepsilon_{\mathbf{k}\sigma} = \varepsilon_{\mathbf{k}} + \sigma^z_{\sigma\sigma} \Delta/2$, where $\sigma^z$ denotes the $z$-component of the Pauli matrices, whereas $\Delta$ defines a rigid shift of the metallic spin-subbands, wee introduce the spin-chemical potentials $\mu_{L\sigma} = \mu_L - \sigma^z_{\sigma\sigma} \Delta/2$

We approach the charge transport properties of the system by considering the flux $I = -e\partial_t \langle N_L \rangle$, which is sufficient in the stationary regime that we are concerned with. Using standard methods we arrive at the expression

$$I = \frac{ie}{h} \sum_\sigma \Gamma^L_\sigma \int \left( f_{L\sigma}(\omega) \mathbb{G}^>_{11\sigma}(\omega) + f_{L\sigma}(-\omega) \mathbb{G}^<_{11\sigma}(\omega) \right) d\omega, \tag{1}$$

where $f_{\chi\sigma}(\omega) = f(\omega - \mu_{\chi\sigma})$ is the Fermi function as the chemical potential $\mu_{\chi\sigma}$, the coupling $\Gamma^\chi_\sigma = 2\pi \sum_{\mathbf{k}} T^2_{\chi\sigma} \delta(\omega - \varepsilon_{\mathbf{k}})$, and $\mathbb{G}^{</>}_{1\sigma}(\omega)$ denotes the lesser/greater forms of the Green function projected onto the molecular level adjacent to the left lead.

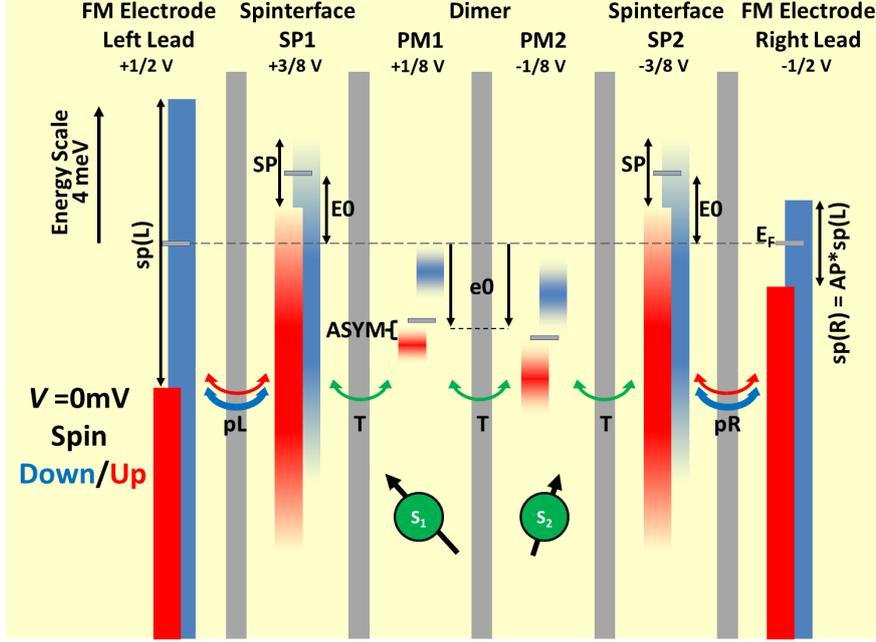

Fig.S 6: **Schematic of the nanotransport path comprising a paramagnetic dimer sandwiched between two outer quantum dots serving as spinterfaces.** $S_1$ and $S_2$ are the spins of the inner quantum dots representing the paramagnetic dimer. The parameters are described in Methods.

*Molecular assembly* - Next, we derive an expression for the Green function of the molecular assembly, in which we include the couplings to the leads as well as the presence of the localized spins. As we have in mind a structure comprising four molecular entities, we shall restrict our calculations to this scenario. We obtain per spin projection the Green function

$$\mathbb{G}_\sigma^{r/a}(\omega) = \left[ \omega - \tilde{\varepsilon}_\sigma - \begin{pmatrix} \mp i \Gamma_\sigma^L/2 & T_1 & 0 & 0 \\ T_1 & 0 & T_2 & 0 \\ 0 & T_2 & 0 & T_3 \\ 0 & 0 & T_3 & \mp i \Gamma_\sigma^R/2 \end{pmatrix} \right]^{-1} \tag{2}$$

under the assumption spin preserving tunneling processes. Here, $\tilde{\varepsilon}_\sigma = \text{diag}\{\tilde{\varepsilon}_{m\sigma}\}_m$ with $\tilde{\varepsilon}_{m\sigma} = \varepsilon_{m\sigma} + v_m \sigma_{\sigma\sigma}^z \langle S_m^z \rangle$. Thanks to this structure of the Green function, we can calculate the lesser/greater forms according to $\mathbb{G}_\sigma^{</>}(\omega) = \mathbb{G}_\sigma^r(\omega) \mathbf{\Sigma}_\sigma^{</>}(\omega) \mathbb{G}_\sigma^a(\omega)$, where all elements vanish except $\Sigma_{11\sigma}^{</>}(\omega) = (\pm i) f_{L\sigma}(\pm\omega) \Gamma_\sigma^L$ and $\Sigma_{44\sigma}^{</>}(\omega) = (\pm i) f_{R\sigma}(\pm\omega) \Gamma_\sigma^R$. It is, then, straightforward to see that

$$\mathbb{G}_{11\sigma}^{</>}(\omega) = (\pm i) \left( f_{L\sigma}(\pm\omega) \Gamma_\sigma^L |\mathbb{G}_{11\sigma}^r|^2 + f_{R\sigma}(\pm\omega) \Gamma_\sigma^R |\mathbb{G}_{14\sigma}^r|^2 \right), \tag{3}$$

which leads to the current

$$I = \frac{e}{h} \sum_\sigma \Gamma_\sigma^L \Gamma_\sigma^R \int \left( f_{L\sigma}(\omega) - f_{R\sigma}(\omega) \right) |\mathbb{G}_{14\sigma}^r|^2 d\omega. \tag{4}$$

By a straightforward calculation we find that

$$\mathbb{G}_{14\sigma}^r(\omega) = -\frac{T_1 T_2 T_3}{\left[(\omega - \tilde{\varepsilon}_{1\sigma} \pm i\Gamma_\sigma^L/2)(\omega - \tilde{\varepsilon}_{2\sigma}) - T_1^2\right]\left[(\omega - \tilde{\varepsilon}_{3\sigma})(\omega - \tilde{\varepsilon}_{4\sigma} \pm i\Gamma_\sigma^R/2) - T_3^2\right] - \alpha} \quad (5)$$

$$\text{with } \alpha = T_2^2(\omega - \tilde{\varepsilon}_{1\sigma} \pm i\Gamma_\sigma^L/2)(\omega - \tilde{\varepsilon}_{4\sigma} \pm i\Gamma_\sigma^R/2)$$

***Exchange interaction*** - We are mainly interested in the variations of the interactions between the two central spins, $\mathbf{S}_2$ and $\mathbf{S}_3$. The outer spins, $\mathbf{S}_1$ and $\mathbf{S}_4$, are not going to be considered since these are expected to be more randomized by the strong coupling of their corresponding molecular states and the states in the leads. The expressions for the non-equilibrium interactions can be found in Ref. [1]. We notice that since the metallic leads are magnetic (Co), it is also necessary to consider the Ising and Dzyaloshinskii-Moriya interactions in addition to the Heisenberg term.

***Thermal protection*** - The set-up with four quantum dots in series leads to an effective thermal protection of the inner quantum dots since the broadening of their levels remains small despite the presence of the leads. It can be viewed as the outer quantum dots partly shielding off the inner ones from the electrodes. This can be understood from two observations. First, a single localized level coupled to a continuum acquires a level broadening due to this coupling. The Green function $G(z) = \langle\langle d_0|d_0^\dagger\rangle\rangle$ for the single level at the energy $\varepsilon_0$ is given by

$$G(z) = \frac{1}{z - \varepsilon_0 - \sum_{\mathbf{k}} |T|^2 g_{\mathbf{k}}(z)}, \quad (6)$$

where $\sum_{\mathbf{k}} |T|^2 g_{\mathbf{k}}(z)$ reflects the tunneling interactions between the localized and de-localized electrons. Defining $\Gamma = -2\mathrm{Im}\sum_{\mathbf{k}} |T|^2 g_{\mathbf{k}}^r(\omega)$ and $E_0 = \varepsilon_0 + \mathrm{Re}\sum_{\mathbf{k}} |T|^2 g_{\mathbf{k}}^r(\omega)$, we can write the retarded Green function as

$$G^r(\omega) = \frac{1}{\omega - E_0 + i\Gamma/2}. \quad (7)$$

The localized level experiences a broadening $\Gamma/2$ due to these interactions.

In the series of quantum dots where only the first and last are coupled to electron reservoirs, the broadening of the inner quantum dot levels become limited by the coupling strength between the quantum dots. For instance, the poles of the four quantum dot system as per above are found by equating the denominator of, e.g., Eq. (5) to zero. Taking, for simplicity $\tilde{\varepsilon}_{1\sigma} = \tilde{\varepsilon}_{4\sigma} = E$, $\tilde{\varepsilon}_{2\sigma} = \tilde{\varepsilon}_{3\sigma} = \varepsilon$, $T_m = T_0$, $m = 1, 2, 3$, and $\Gamma_\sigma^\chi = \Gamma$, this equation for the retarded version reduces to

$$\left[(\omega - E + i\Gamma/2)(\omega - \varepsilon) - T_0^2\right]^2 - T_0^2(\omega - E + i\Gamma/2)^2 = 0, \quad (8)$$

with the solutions

$$\varepsilon_\pm = \frac{1}{2}\left(E + \varepsilon + T_0 - i\frac{\Gamma}{2} \pm \sqrt{\left(E - \varepsilon - T_0 - i\frac{\Gamma}{2}\right)^2 + 4T_0^2}\right) \quad (9a)$$

$$E_\pm = \frac{1}{2}\left(E + \varepsilon - T_0 - i\frac{\Gamma}{2} \pm \sqrt{\left(E - \varepsilon + T_0 - i\frac{\Gamma}{2}\right)^2 + 4T_0^2}\right) \quad (9b)$$

The limit of our interest is given for $\Gamma/2 \gg 2T_0$, which gives the approximate solutions

$$\varepsilon_\pm \approx \begin{cases} E - i\frac{\Gamma}{2}, & (+) \\ \varepsilon + T_0 - i\frac{\Gamma}{2}\frac{T_0^2}{(E - \varepsilon - T_0)^2 + (\Gamma/2)^2}, & (-) \end{cases} \quad (10a)$$

$$E_\pm \approx \begin{cases} E - i\frac{\Gamma}{2}, & (+) \\ \varepsilon - T_0 - i\frac{\Gamma}{2}\frac{T_0^2}{(E - \varepsilon + T_0)^2 + (\Gamma/2)^2}, & (-) \end{cases}. \quad (10b)$$

Among those four poles, the two energies $\varepsilon_+$ and $E_+$ are weighted on the outer quantum dots whereas the other two, $\varepsilon_-$ and $E_-$ are weighted on the inner. These latter energies acquire a level broadening which is maximally $2T_0^2/\Gamma \ll \Gamma/8$, according to the assumption.

***Thermal fluctuations*** - The thermal fluctuations of the charge in the quantum dots can be estimated by calculating the occupation numbers $\langle n_m \rangle$, $m = 1, \ldots, 4$. In terms of the non-equilibrium formalism we use for the calculations in this context, the occupation numbers can be related to the lesser Green functions through the relation $\langle n_m \rangle = (-i) \sum_\sigma \int \mathbb{G}^<_{mm\sigma}(\omega) d\omega / 2\pi$.

First, however, consider the occupation number of the single localized level from the previous section. This is given by

$$\langle n \rangle = \frac{\Gamma}{2\pi} \int \frac{f(\omega)}{(\omega - E_0)^2 + (\Gamma/2)^2} d\omega. \tag{11}$$

Then, in the low temperature limit, the condition $\Gamma/2 \gg k_B T$ holds. In this limit, the occupation number approaches unity (zero) whenever $E_0 - \mu < 0$ ($E_0 - \mu > 0$), something that can be seen from the expression

$$\langle n \rangle \approx \frac{1}{\pi} \left( \arctan \frac{\mu - E_0}{\Gamma/2} + \frac{\pi}{2} \right): \ \frac{\Gamma}{2} \gg k_B T. \tag{12}$$

It can be seen, nonetheless, that the occupation number is independent, or at most weakly dependent, of the temperature in this limit.

In the high temperature limit, on the other hand, the condition $\Gamma/2 \ll k_B T$ is satisfied, which means that the position of the localized level relative to the chemical potential is almost insignificant within the energy range defined by the temperature. This is quantified by

$$\langle n \rangle \approx \frac{1}{2\pi} \left( \arctan \frac{\mu + k_B T/2 - E_0}{\Gamma/2} - \arctan \frac{\mu - k_B T/2 - E_0}{\Gamma/2} \right)$$
$$\approx \frac{1}{2}, \ |\mu - E_0| \ll \frac{k_B T}{2}: \ \ \frac{\Gamma}{2} \ll k_B T. \tag{13}$$

The intermediate regime, in which $\Gamma/2 \sim k_B T$, cannot be quantified in the same simple forms as the high and low temperature limits.

# 6 Spin-, bias- and energy-dependent DOS on the analytical model's four QDs

As explained in the main text we propose the concept of spin engine which is harvesting the spin fluctuations that are induced by thermal fluctuations on the PM center. In this section we present the spin-, bias- and energy dependencies of our analytical model which servers to explain the physics of such a spin engine. We present in Fig.S 7 the spin-resolved energy and bias dependence of the DOS of the analytical model's 4 quantum dots in the MTJ's P and AP magnetic states. Within the scope of the anaylitical model the spin engine is composed of two ferromagnetic electrodes (left/right leads), of two spinterfaces SP1 & SP2 and of a PM dimer PM1 & PM2. As expected, the energy dependence of the spinterfaces SP1 & SP2 tracks the bias-voltage imposed shift in energy position, with little bias voltage-induced anomalies. In contrast, a much more complex picture of voltage-dependent weighing of the spin-polarized bonding and antibonding states appears on the paramagnetic dimer (PM1 & PM2). At V = 0 and in the MTJ's P magnetic state, we observe two branches of DOS: one near $E_F$, and the other at -5meV. On PM1, the top(bottom) branch has a majority spin ↓(↑) states. This is reversed on PM2, meaning that there is a net AF coupling of paramagnetic fluctuations between the two. In the AP magnetic state, the situation becomes more complex.

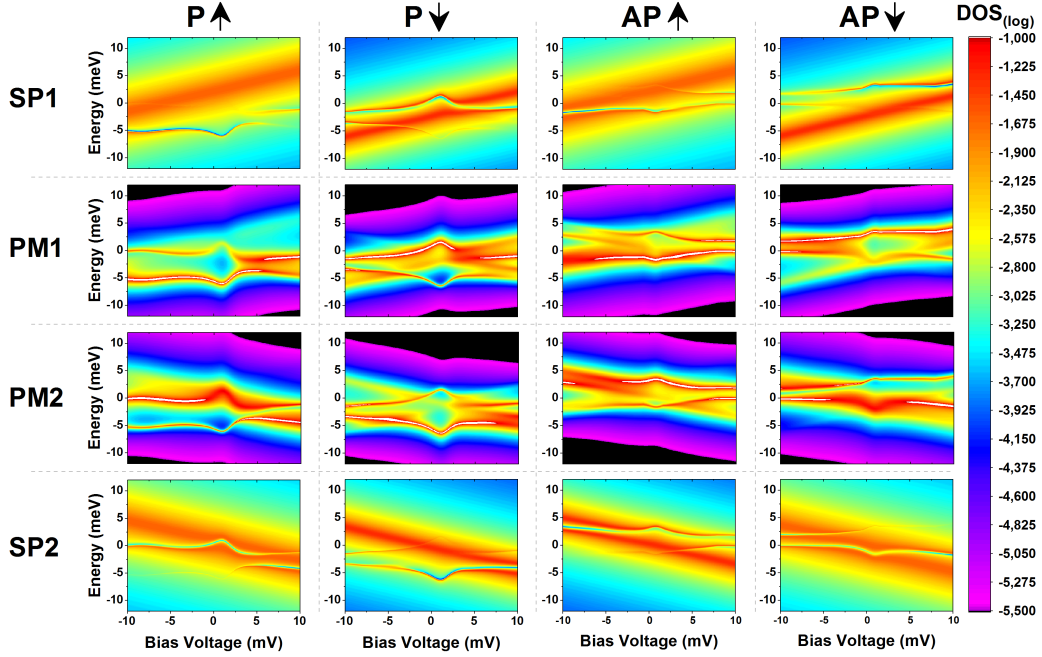

Fig.S 7: Spin-resolved energy and bias dependence of the DOS of the analytical model's 4 quantum dots SP1, PM1, PM2 & SP2 in the MTJ's P and AP magnetic states,. Compared to the case of the spinterfaces SP1 and SP2, the PM dimer, represented as PM1 & PM2, exhibits a complex, bias-dependent energy dependence of spin non-degenerate bonding/antibonding states.

# 7 Impact of C-C distance on effective magnetic coupling

We performed DFT calculations for the C-doped MgO using 64-atom supercells with a simple cubic structure with two substitutional carbon atoms in various configurations (see Fig.S 8). These calculations were done using VASP code [2] based on the projector augmented wave (PAW) method[3] and the Pedrew, Burke, Enzerhof (PBE)[4] generalized gradient approximation for the exchange-correlation potential. The kinetic energy cutoff value of 500 eV for the plane wave basis set and the convergence criterion for the total energy of $10^{-8}$ eV is used. The carbon doped structures are fully relaxed using conjugate-gradient algorithm with requirement that the forces acting on atoms are less than 0.001 eV/Å. A k-point mesh of $6\times6\times6$ with the Methfessel-Paxton method with a smearing of $\tau= 0.1$ eV is used.

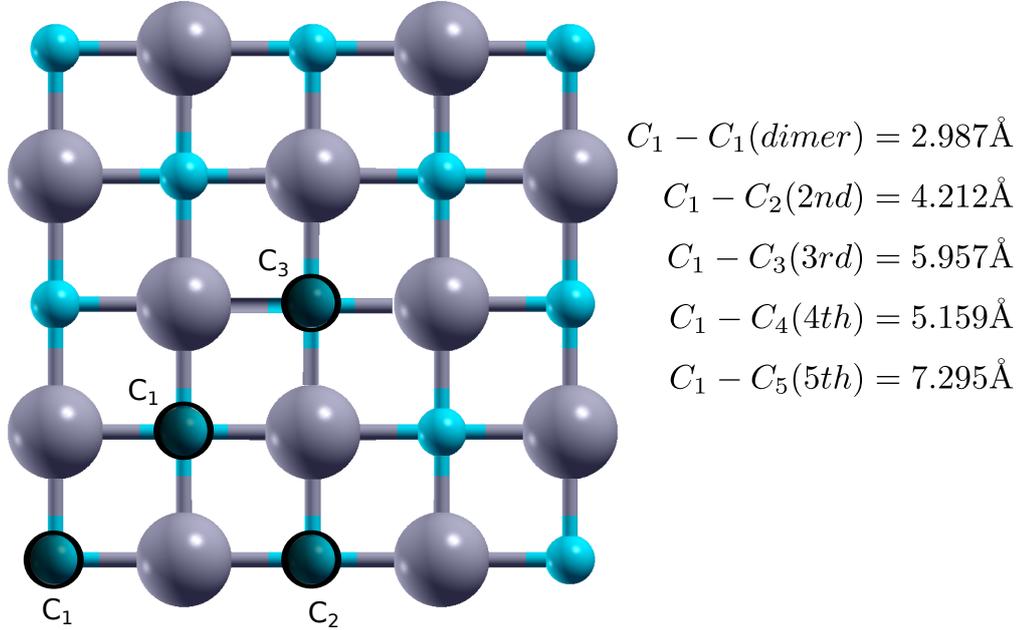

$$C_1 - C_1(dimer) = 2.987 \text{Å}$$
$$C_1 - C_2(2nd) = 4.212 \text{Å}$$
$$C_1 - C_3(3rd) = 5.957 \text{Å}$$
$$C_1 - C_4(4th) = 5.159 \text{Å}$$
$$C_1 - C_5(5th) = 7.295 \text{Å}$$

Fig.S 8: Configuration of C-C atoms in MgO supercell replacing oxygen atoms with distances indicated. The 4th and 5th neighbours are on diagonals. Distances correspond to structures before relaxation.

In agreement with previous theoretical calculations [5], we found strong pairing interaction between C-C atoms with the shortest distance. In fact, relaxation revealed that this distance decreases from initial 2.87 Å to 1.4885 Å and the corresponding total energy is lower at most by 3.02 eV compared to the well-separated C impurities (see Tab. 1).

The coupling between substitutional carbon atoms (dimer) leads to the creation of additional energy levels within MgO band gap (Fig.S 9). These are the bonding and the antibonding states due to pp$\pi$ and pp$\sigma$ levels where the pp$\pi$ antibonding state is half-filled and it is in a spin triplet state (see Ref. [5]).

In the case of a single carbon impurity, we observed fully occupied spin up levels separated from 1/3 filled spin down states via the exchange splitting interaction (Fig.S 9 left). As a consequence single carbon substitution can also lead to creation of paramagnetic state in MgO.

| Structure | $\mu_{FM}$ | FM [eV] | $\mu_{AF}$ | AF [eV] | $\Delta J$(AF-FM) |
|-----------|-----------|---------|-----------|---------|------------------|
| 1xC | 0.725 | -375.14200273 | - | - | - |
| 2xC(dimer) | 0.336 | -372.30848248 | 0.0 | -372.05708927 | 0.2514 |
| 2xC(2nd) | 0.721 | -369.49545665 | ± 0.720 | -369.37657882 | 0.1189 |
| 2xC(3rd) | 0.739 | -369.64105037 | ± 0.700 | -369.55435115 | 0.0867 |
| 2xC(4th) | 0.727 | -369.5255653 | ±0.728 | -369.65011625 | -0.1246 |
| 2xC(5th) | 0.725 | -369.2877705 | ±0.724 | -369.27779311 | 0.00998 |

Table 1: Total energy and magnetic moments for C in C-MgO and Si in Si-MgO structures after relaxation.

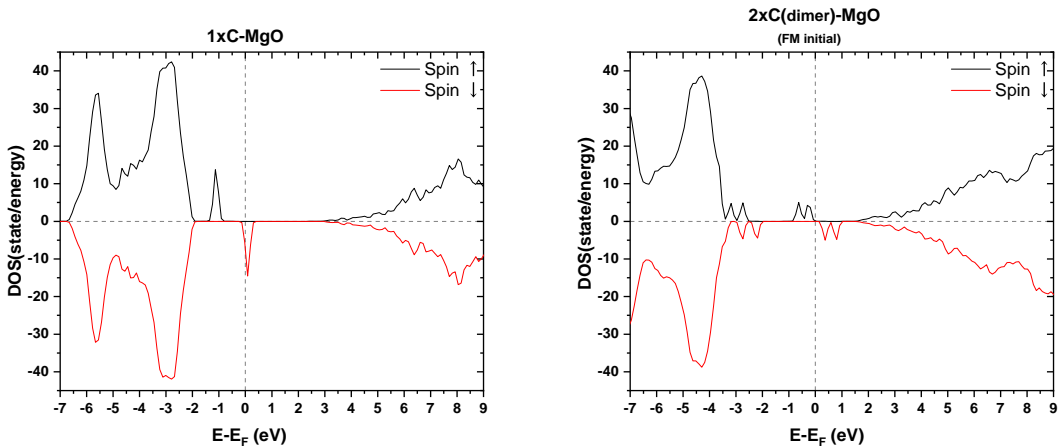

Fig.S 9: Density of states (DOS) for MgO bulk doped with one (left panel) and two (right panel) carbon atoms ("dimer").

# 8  Other origins of the paramagnetic centers within the barrier

As discussed in the main text, the MgO of our MTJs contains single and double oxygen vacancies. To examine whether these vacancies can capture C atoms that diffuse during annealing, we calculated formation energy of carbon monomer/dimer and of single/double oxyge vacancy (F/M center respectively) :

$$E_f[X] = E_{tot}[X] - E_{tot}[bulk] - \sum_i n_i \mu_i, \qquad (14)$$

where $E_{tot}[X]$ is the total energy of a supercell with a defect $X$, and $E_{tot}[bulk]$ is the total energy for the ideal structure using the same supercell. The number $n_i$ corresponds to the number of atoms of type $i$ which are added ($n_i > 0$) or removed ($n_i < 0$) from the supercell, and $\mu_i$ denotes the chemical potential of a particular species. To mimick experimental conditions, for the formation energy of carbon impurities we used as a referential ($E_{tot}[bulk]$) MgO cell with a F/M center generated in it, while for the formation energy of F/M center, we used an ideal MgO supercell. Our results are summarized in the table. We find that carbon capture by single/double oxygen vacancies, which can be present in MgO, is energetically favorable and can yield both paramagnetic monomers and dimers.

This shows that C monomers are in principle possible. To test whether the quantum object in the MTJ's nanotransport path could be a paramagnetic monomer, we adapted our analytical model to this case. We then explored the model's parameter space under similar physical constraints as those used in the dimer calculation (see Main text and Section. 5): - a reasonable correspondence between the spin polarization amplitude of the two FM electrodes. - the parameters describing the electrodes and spinterfaces should be the same for the MTJ's P and AP magnetic states. - only a small variation in the parameter e0 describing the energy position of the monomer is possible.

| Structure | Formation energy [eV] |
| --- | --- |
| 1xC | -5.843 |
| 2xC(dimer) | -14.5936(FM) / -14.3422(AF) |
| F-MgO | 7.3432 |
| M-MgO | 14.5871 |

Table 2: Formation energies of C-MgO, F-MgO and M-MgO. The reference energy $E_{tot}[bulk]$ corresponds to that of F/M-MgO supercells.

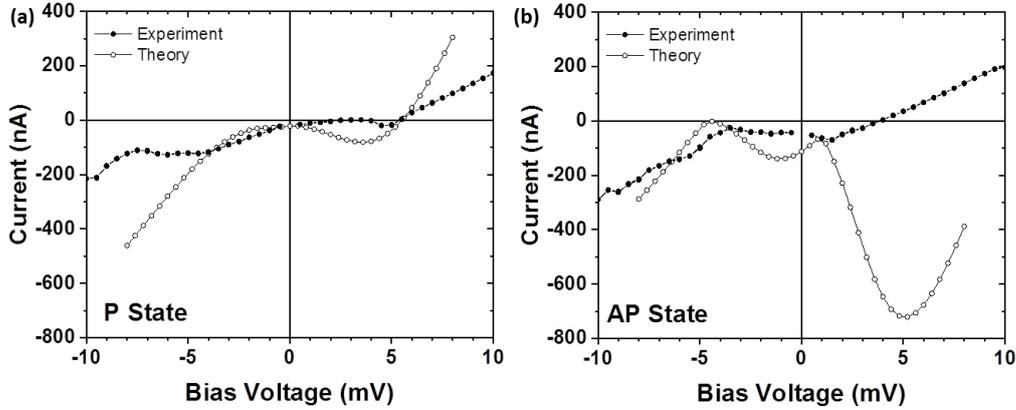

Fig.S 10: **Attempted reproduction of experimental data using a monomer-based model** for the MTJ's (a) P and (b) AP magnetic states. Parameter values are: sp = 11, SP1=2, pR=pL=0.3, AP=0.1, SP2=1, E0=0, and for the P(AP) state eo=0.2(-0.1), all in meV. While the P fit is moderately good, the corresponding AP fit is very poor.

We began with parameter values similar to those used in the dimer calculation. We present in Fig.S 10 an illustrative result. Panel (a) shows how parameter space may be tuned in order to reproduce, with a fidelity comparable to the dimer case, the experimental I(V) data in the MTJ's P magnetic state. However, it is then impossible to reproduce the experimental I(V) data for the MTJ's AP magnetic state (see panel b). We attribute this result to the presence of only one parameter describing the quantum object here, as opposed to the two that described the paramagnetic dimer. We therefore conclude that the quantum object in the nanotransport path cannot be a paramagnetic monomer.

Finally, although charged single oxygen vacancies are paramagnetic [6], and can in principle form a paramagnetic dimer through proximity-induced magnetic coupling, this scenario is inconsistent with our analytical theoretical results. Indeed, the localized state of neutral single/double oxygen vacancies already lies at least 0.2eV below EF for a Co electrode[7], and charging these levels lowers their energy due to electrostatics by 0.5eV[8]. Charged oxygen vacancies thus cannot explain the analytical model's spin-split states at $E_F$.

# 9 Determination of the $E_F$ energy level in the main text's Fig. 4

The main text's Fig. 4 shows the DOS of MgO containing a pair of C atoms in 4th nearest-neighbor configuration with an AF coupling. To determine whether this system would exhibit states at $E_F$ in a Co/MgO/Co MTJ, we computed the electronic properties of the C dimer within the MgO barrier of the Co/MgO/Co trilayer using the same formalism. A kinetic energy cutoff value of 500 eV for the plane wave basis set and a convergence criterion for the total energy of $10^{-7}$eV were used. A k-point mesh of $4\times4\times1$ with the Methfessel-Paxton method with a smearing of $\tau = 0.1$ eV was found to be sufficient to properly describe electronic structure of the system. For simplicity, the geometry of the MgO/Co supercell's geometry follows that of MgO/Fe, with the lateral lattice constant of MgO and the transverse direction scaled accordingly. Oxygen atoms were placed on top of Co atoms and the Co-O distance at the interface was fixed to 2.17Å.

Ideally, we would compute these properties for a MTJ stack comprising the 2.5nm (i.e. 12 ML) experimental thickness of MgO. However, this system is too large, which generates energy convergence problems. We therefore computed for thinner (6ML, 8ML) layers of MgO, using 4-ML-thick Co electrodes. We find that, for 6ML (8ML), the MTJ's Fermi level lies 2.68(2.57)eV above the valence band edge (data not shown). We therefore extrapolated the value 2.35eV for this energy distance in the case of 12ML, which we used to peg $E_F$ in the main text's Fig. 4.

# 10 Information-to-energy mechanism

We presented in the main text a spin engine concept based on maintainin the magnetic state of the PM nanoobject and FM electrodes. We experimentally observe the rectification of thermal fluctuations as a signature of this spin engine's operation. Our analytical and ab-initio theories support linking our experimental observation of electrical power generation to our spin engine concept. This is in particular supported by the quality of the agreement between our analytical theory and experiment.

To expand the discussion, we noted in the main text some obvious links between our work and that of quantum thermodynamics. We mentioned a Maxwell demon involving spin-polarized transport across two magnetically coupled centers, as reported by Ptaszyński [9]. Therein, control over locally negative information entropy leading to extractable work arises from fully spin-polarized transport at thermal equilibrium toward one of two magnetically coupled QDs (which would correspond to PM1 & PM2 in our analytical model). Therein, each dot provides spin-based feedback control on the other's effective spin-polarized tunnelling rates by reading out its spin state through the magnetic coupling. As a note, the nanotransport path described therein proceeds in parallel along the two quantum-coupled centers, while it is in series in our analytical model.

The spin engine contains two ingredients of an information-to-energy conversion process [10, 11]: a rectification of thermal fluctuations, and an increase in information entropy. Since the former has already been described, we mention here that information entropy can increase along our path if an electron leaving the spinterface with a well-determined spin (not only by virtue of a spin referential that is fixed by the adjacent FM electrode, but also due to its total spin polarization) arrives onto a PM center. Indeed, with evolving time, this electron's spin is no longer determined. Note that this would occur only if the system experiences quantum decoherence. This increase in information entropy would then be partly reutilized to perform work. We surmise that information erasure arises when the electron leaves the PM center and is 'read' out by a spinterface, as explained by Ptaszyński [9].

## 11 Calculation of power generation under resistive load & comparison with solar cells

The MTJ generates a power P>0.1nW in its AP magnetic state within $+1.4 < V(mV) < +2.4$, i.e. for an output resistance range $20 < R(k\Omega) < 60$. P>0.1nW in the P magnetic state occurs for R=200k$\Omega$. The following discussion is intended to provide a figure of merit for the concept's potential compared to other renewable energy generation sources. The discussion assumes that reproducibility has been achieved, and that these power MTJs can be routinely designed into arrays as are their ICT MTJ counterparts in MRAM technologies. We furthermore assume that the 0.1nW measured on our 20$\mu m$-wide MTJ would also apply to a sub-micronic junction. This is reasonable considering that the device effectively operates across a nanotransport path, but reducing the device size might exacerbate heating-induced degradation of the MTJ's magnetic state during operation, and thus performance.

A 2016 MRAM announcement [12] by SD Hynix and Toshiba shows a square array of MTJs with a pitch of 90nm. Extrapolating to 1 cm$^2$ yields $\approx 1.2 \times 10^{12}$ devices. At this density, and assuming a load resistance equal to the power source's output resistance, with a 0.1nW output power per MTJ, we find an output power on the load of 58.8W, i.e. an available power density of 58.8W/cm$^2$.

A series/parallel assembly of MTJs may be implemented to tune the output resistance so as to ensure optimal power transfer to a load resistance of equal value. As an example, we use R=20k$\Omega$ and operating voltage $V$=1.4mV from our experimental data, and connect the $1.2 \times 10^{12}$ MTJs in the following series/parallel geometry. We put 2500 MTJs in series, thereby yielding 3.5V output voltage and an output resistance R=48M$\Omega$. This series circuit is then placed in parallel $4.8 \times 10^8$ times. The resulting source has a bias of 3.5V and an output resistance of 0.1$\Omega$. With an equal 0.1$\Omega$ load resistance, this bias drop generates 58.8W on the load, and 58.8W on the source's output resistance. Other series/parallel combinations can be used to increase/decrease the bias drop and output resistance simultaneously while delivering the same 58.8W to a balanced load.

This calculation supposes the heat dissipated by the source's output resistance is properly evacuated, and the absence of bias drop on interconnects. Suppose that, for these and other possible reasons, the output power is only 1% of the nominal output power. The resulting $\approx$6000W/m$^2$ power density, calculated based on our unoptimized device, would nevertheless still be 6x greater than the raw power density due to solar irradiation (1000W/m$^2$ in standard (terrestrial, temperate) temperature and conditions (STC) [13], which is then lowered by the solar cell's efficiency[14]. As a guideline, present-day commercial solar cells have an efficiency of $\approx$20%). This discussion suggests that, once reproducibility is achieved in interconnected, arrayed devices, the power MTJ concept could strongly alter the renewable energy mix.

# References

[1] J. Ren J. Fransson and J. X. Zhu. Efficient iterative schemes for ab initio total-energy calculations using a plane-wave basis set. *Physical Review Letters*, 113(25):257201, 2014.

[2] G. Kresse and J. Furthmüller. Efficient iterative schemes for ab initio total-energy calculations using a plane-wave basis set. *Physical Review B*, 54(16):11169, 1996.

[3] G. Kresse and D. Joubert. From ultrasoft pseudopotentials to the projector augmentedwave method. *Physical Review B*, 59(3):1758, 1999.

[4] J. P. Perdew, K. Burke, and M. Ernzerhof. Generalized gradient approximation made simple. *Physical Review Letters*, 77(18):3865, 1996.

[5] H. Wu, A. Stroppa, S. Sakong, S. Picozzi, M. Scheffler, and P. Kratzer. Magnetism in c- or n-doped mgo and zno: A density-functional study of impurity pairs. *Physical Review Letters*, 105(26):267203, 2010.

[6] K. P. McKenna and J. Blumberger. Crossover from incoherent to coherent electron tunneling between defects in MgO. *Physical Review B*, 86(24):1–5, 2012.

[7] F. Schleicher *et al.* Explicit proof of hole transport across a MgO-based magnetic tunnel junction due to oxygen vacancies. *Manuscript submitted for publication, arXiv:1711.05643v1*, 2018.

[8] K. P. McKenna and A. L. Shluger. First-principles calculations of defects near a grain boundary in MgO. *Physical Review B*, 79(22), 2009.

[9] K. Ptaszyński. Autonomous quantum maxwell's demon based on two exchange-coupled quantum dots. *Physical Review E*, 97(1):012116–8, 2018.

[10] D. Mandal and C. Jarzynski. Work and information processing in a solvable model of maxwell's demon. *Proceedings of the National Academy of Sciences*, 109(29):11641–11645, 2012.

[11] P. Strasberg, G. Schaller, T. Brandes, and M. Esposito. Quantum and information thermodynamics: A unifying framework based on repeated interactions. *Physical Review X*, 7(2):021003–33, 2017.

[12] S. W. Chung *et al.* 4gbit density STT-MRAM using perpendicular MTJ realized with compact cell structure. In *2016 IEEE International Electron Devices Meeting (IEDM)*, pages 27.1.1–27.1.4. IEEE, 2016.

[13] S. Indrajit, S. Mukherjee, and K. Biswas. A review of energy harvesting technology and its potential applications. *Environmental and earth sciences research journal*, 4(2):33–38, 2017.

[14] T. G. Walmsley, M. R.W. Walmsley, P. S. Varbanov, and J. J. Klemeš. Energy ratio analysis and accounting for renewable and non-renewable electricity generation: A review. *Renewable and Sustainable Energy Reviews*, 98:328–345, 2018.